%% file: main.tex
\documentclass[sigconf]{acmart}
\usepackage{amsmath,amsfonts}
\usepackage{algorithmic}
\usepackage{graphicx}
\usepackage{textcomp}
\usepackage{xcolor}
\usepackage[ruled, vlined,linesnumbered]{algorithm2e}
\usepackage{epstopdf}
\usepackage{comment}
\usepackage{multirow}
\usepackage{url}
\usepackage{optidef}
\usepackage{subcaption}
\usepackage{listings}
\usepackage{amsthm}
\usepackage{subcaption}
\theoremstyle{definition}

\newcommand{\our}{\textit{VDMS-Async}\xspace}

\AtBeginDocument{%
  \providecommand\BibTeX{{%
    \normalfont B\kern-0.5em{\scshape i\kern-0.25em b}\kern-0.8em\TeX}}}

\setcopyright{acmcopyright}
\copyrightyear{2018}
\acmYear{2018}
\acmDOI{XXXXXXX.XXXXXXX}

\acmConference[Conference acronym 'XX]{Make sure to enter the correct
  conference title from your rights confirmation emai}{June 03--05,
  2018}{Woodstock, NY}
\acmPrice{15.00}
\acmISBN{978-1-4503-XXXX-X/18/06}

\begin{document}

\title[\our]{Towards a Flexible Scale-out Framework for Efficient\\ Visual Data Query Processing}

\author{Rohit Verma}
\email{rohit1.verma@intel.com}
\affiliation{%
  \institution{Intel Labs}
  \country{India}
}

\author{Arun Raghunath}
\email{arun.raghunath@intel.com}
\affiliation{%
  \institution{Intel Labs}
  \country{USA}
  }








\begin{abstract}
There is growing interest in visual data management systems that support queries with specialized operations ranging from resizing an image to running complex machine learning models. With a plethora of such operations, the basic need to receive query responses in minimal time takes a hit, especially when the client desires to run multiple such operations in a single query. Existing systems provide an ad-hoc approach where different solutions are clubbed together to provide an end-to-end visual data management system. Unlike such solutions, the Visual Data Management System (VDMS)~\cite{remis2021using} natively executes queries with multiple operations, thus providing an end-to-end solution. However, a fixed subset of native operations and a synchronous threading architecture limit its generality and scalability. 

In this paper, we develop \our that adds the capability to run user-defined operations with VDMS and execute operations within a query on a remote server. \our utilizes an event-driven architecture to create an efficient pipeline for executing operations within a query. Our experiments have shown that \our reduces the query execution time by $2-3$X compared to existing state-of-the-art systems. Further, remote operations coupled with an event-driven architecture enables \our to scale query execution time linearly with the addition of every new remote server. We demonstrate a 64X reduction in query execution time when adding 64 remote servers.
\end{abstract}





\maketitle

\input{Introduction}
\input{Background}
\input{VDMS}
\input{RemoteOperation.tex}
\input{AsyncModel}
\input{Evaluation.tex}
\input{Conclusion}


\bibliographystyle{ACM-Reference-Format}
\bibliography{ref}

\end{document}

%% file: Introduction.tex
\section{Introduction}\label{introduction}
Streaming, virtual reality, IoT, and Smart City becoming household terms has re-ignited the interest in visual data management research~\cite{zhao2022analysis,sensortowerstreaming, vrtrend}. Visual data applications require several data processing tasks (filtering, merging, cropping, etc.) as well as specialized operations (resizing, AI/ML inference like object detection, etc.). All these tasks have to be performed quickly for a better user experience. Here, emphasis must be put on the term "\textit{quickly}", tied to the query execution time, which is a crucial metric linked to user experience. A \textit{quicker} execution time would be possible through efficient visual data management and intelligent query processing. However, several challenges arise when addressing the problem of optimal query execution time.

The first challenge with visual data management is linked to applications executing multiple operations in a pipeline. In the absence of an end-to-end system, these applications work with multiple solutions, each performing a different task. For example, an image classification application will require (i)~creating the image metadata based on the features required by the classification model, (ii)~filtering the data based on constraints, (iii)~pre-processing the data based on the model requirements, and (iv)~executing the model on the images to classify. This pipeline would require (i)~solution to extract and label necessary metadata (feature engineering tool), (ii)~metadata filtering solution (like MySQL~\cite{mysql2001mysql}), (iii)~pre-processing tool, and (iv)~client to execute the classification model. Thus, four different solutions are required to perform the classification pipeline instead of running it from a single DBMS like MySQL. Stitching together multiple disparate solutions is difficult to manage, often error-prone, and requires a separate programming lifecycle for each solution. Moreover, moving large amounts of data back and forth between these distinct processes is inefficient. 

The second challenge arises when operations are compute-intensive enough to require a long processing time and thus stall the overall query execution. 

Availability of resources poses the third challenge, especially in resource-constrained environments where the visual data management systems might not have sufficient compute to perform the compute-intensive operations in a timely manner. Further, such systems often have a limited available storage footprint. Hence, utilizing capable remote servers to assist with the computing and data storage tasks becomes an important consideration. Offloading further adds to the challenges of handling low network bandwidth, communication failure, and stalling. 


A majority of the existing solutions work with ad-hoc application designs that implement different operations using disparate/distinct frameworks. For example, Facebook utilizes Tao~\cite{venkataramani2012tao}, Haystack~\cite{beaver2010finding}, and f4~\cite{muralidhar2014f4}, alongwith MySQL for their photo management applications. In contrast to such ad-hoc solutions, VDMS~\cite{remis2021using} provides an end-to-end solution that stores and manages both visual data and related metadata. However, VDMS does not support user-defined operations or remote computation of operations. Moreover, it follows a synchronous model to execute an operation pipeline that could lead to execution delays (Details in Section~\ref{vdms}). Other existing video data management systems~\cite{poms2018scanner,haynes2018lightdb,kang2017noscope} are also synchronous or use a restricted parallelization approach. Furthermore, these solutions do not support operation execution at a remote server and require an ad-hoc setup to achieve remote execution.

In light of the limitations of the existing systems, we have developed an improved version of VDMS ($\our$) that provides an end-to-end solution to address the issues linked to compute-intensive operation pipelines, limited resource availability, and synchronous query execution. The key contributions are:
\begin{itemize}
    \item A design that allows adding user-defined operations (UDF) as part of a query pipeline. This capability expands the reach of VDMS to run any type of operation over its data store.
    \item A flexible query interface providing a plug-and-play mechanism for executing UDF on remote servers with no code changes to VDMS.
    \item An asynchronous event-driven architecture for visual query processing that avoids stalls in the query pipeline due to communication delays or compute-intensive operations. \our is the first system in literature to utilize an asynchronous event-driven architecture for visual data management resulting in lower query duration and higher throughput than its counterparts.
    \item Detailed benchmarking and analysis to highlight the scale-out benefits of remoting coupled with an asynchronous threading  model by using an ecosystem of remote servers
\end{itemize}

Our experiments performed on image and video datasets show that \our decreases the query execution time by $2-3$X when compared to existing systems for both simple and compute-intensive queries. \our shows $~3$X reduction in execution time for both video and images when multiple parallel clients (as high as 128 parallel clients) perform compute-intensive queries. Furthermore, scaling out to additional servers decreases the execution time linearly with the addition of new remote servers. Our experiments show that the query execution time is reduced by $64$X when 64 remote servers are used.

In the next section, we provide an overview of the existing systems and their limitations (\S~\ref{background}), followed by an overview of VDMS (\S~\ref{vdms}) in the following section. We then discuss the design of remote operations functionality of VDMS along with user-defined operations (\S~\ref{remote}). The following section describes the asynchronous execution model (\S~\ref{asyncmodel}). We evaluate \our over video and image datasets and compare its performance with respect to existing systems (\S~\ref{evaluation}). We then finally conclude in \S~\ref{conclusion}.

%% file: Background.tex
\section{Related Work}\label{background}
Works like Chabot~\cite{ogle1995chabot} and QBIC~\cite{flickner1995query} led the foundation of research in the area of visual data management. These were followed by a large set of visual data management systems that were designed to address different aspects of visual data, such as efficient data store (Strg-index~\cite{lee2005strg}), intelligent query systems (Jacob~\cite{la1996jacob}), efficient metadata management (Oh et al.~\cite{oh2000efficient}), and efficient data mining for application design (Zhu et al.~\cite{zhu2005video}). However, these systems addressed different disjoint aspects and especially are not designed to work with compute-intensive operations leading to the need for efficient visual data management systems that would club all above mentioned aspects of visual data management. 

A majority of the current end-to-end systems rely on ad-hoc solutions that require stitching together multiple disparate tools to create an end-to-end application. Furthermore, most of these solutions either cater to only videos or only images. Facebook created their photo management application as an ad-hoc solution of Tao~\cite{venkataramani2012tao} (data store), Haystack~\cite{beaver2010finding} (object store), f4~\cite{muralidhar2014f4} (blob storage system), and MySQL~\cite{mysql2001mysql} (metadata storage). Many solutions have been developed recently to cater to video data specifically~\cite{daum2021tasm, bastani2020vaas, lai2021top, kang2018blazeit, poms2018scanner, haynes2018lightdb, kang2017noscope}. Scanner~\cite{poms2018scanner} is an open-source system designed for efficient video processing. It provides tools and APIs to perform a variety of video processing operations such as object detection, face detection, pose detection, etc. LightDB~\cite{haynes2018lightdb} is another video-only solution designed especially for virtual reality videos at scale. It, too, provides support to execute specialized operations on the data store. However, LightDB works with angles instead of pixels, and thus all operations have to map the two coordinate systems, which could lead to complications. NoScope~\cite{kang2017noscope} is a specialized video data management system designed to support deep learning operations at scale. But, it does not support any operations that do not require deep learning models. 
Moreover, these solutions assume the availability of capable servers that can perform expensive computing tasks locally.

Current iterations of MySQL~\cite{mysql2001mysql} support storing images and videos as binary data. It is also possible to execute user-defined operations using SQL functions, but most specialized operations, such as executing machine learning classification tasks, need to be performed using an external client, requiring moving data back and forth, which is inefficient. PostgreSQL~\cite{momjian2001postgresql} also provides a similar storage solution with a much broader scope for executing specialized operations using Python libraries. However, parallelization of these operations is limited by the number of available CPUs and might lead to stalling when working with a large dataset or long-running compute-intensive operations.

Several works have tried to optimize query execution time using different techniques~\cite{rheinlander2015sofa, ramachandra2017froid, jahani2011automatic, hellerstein1993predicate}. Recent techniques like LPCE~\cite{wang2023speeding} utilize machine learning techniques for cardinality estimation to decrease the overall query execution time. However, these solutions either need to know what type of query is being performed or historical query information, rendering them less effective for queries with user-defined operations (UDF). Works like Voodoo~\cite{he2020method} provide a method to optimize query execution time for UDFs using indexing but are limited to images only. 

%% file: VDMS.tex
\section{VDMS Overview}\label{vdms}

The Visual Data Management System (VDMS)~\cite{remis2021using} was designed for swift visual data access by influencing how the data is treated in the storage system. VDMS uses an in-memory persistent graph data store~\cite{gupta2017addressing} to enable fast meta-data access. VDMS runs as a server listening for client queries for data management.

\begin{figure}[!ht]
    \centering
    \includegraphics[width=0.4\linewidth]{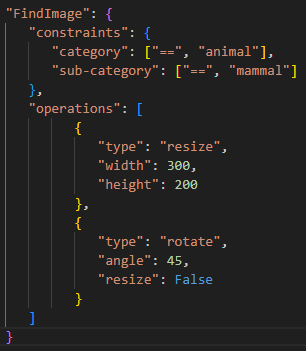}
    \caption{VDMS query with operation pipeline}
    \label{fig:query-example}
\end{figure}

VDMS provides the capability to run specialized functions on visual data. These operations can be executed while adding the visual data or querying the database. VDMS natively provides a small set of OpenCV~\cite{bradski2000opencv} operations, such as crop, trim, rotate, etc. All VDMS operations can be executed in a pipeline. For example, in Figure~\ref{fig:query-example}, an image search query is run that also requests a resize, followed by a rotate operation on the returned images matching the query. VDMS ensures that the operations are executed on all the images that match the search criteria in the order specified by the user. Further, VDMS also provides specific queries on metadata that could be used to perform analytics, such as similarity search.

The prime advantage of VDMS versus other visual data management systems is the capability to provide end-to-end support for performing visual data queries. However, there exist some key limitations of the current version of VDMS. First, VDMS does not have the support to execute user-defined operations on the visual data. Although the native operations are suitable for a large subset of visual data queries~\cite{remis2021using}, the absence of support for user-defined operations restricts the type of operations required for real-world applications, as discussed in Section~\ref{introduction}. Second, VDMS does not support executing operations on a remote server, which might be a crucial requirement for different applications (Section~\ref{introduction}). Finally, VDMS executes the operation pipeline synchronously. Hence, if it takes $t$ seconds to execute all operations on one image, for $10$ images, the total query execution time will be $\approx (t*10)$ seconds.



%% file: RemoteOperation.tex
\section{User-Defined and Remote Operations} \label{remote}

In this section, we describe how we extend VDMS to support user-defined and remote operations. User-defined operation adds the capability to run specialized operations designed by a user as part of the VDMS operation pipeline without requiring any manual stitching. Remote operations add the capability to offload operations to a remote server from VDMS such that the VDMS server can continue working on other tasks.

\subsection{User-Defined Operations}
As discussed in the previous section, VDMS only provides support to run basic OpenCV operations on the visual data, which severely limits the reach of VDMS. Although the existing set of operations can be used to perform a large set of visual data queries in comparison to existing state-of-the-art solutions~\cite{remis2021using}, complex tasks like running machine learning models or complex visual data processing would require an ad-hoc solution. Such ad-hoc solutions would query VDMS to perform some basic operations, retrieve all the visual data responses and then execute the complex operations in a separate process. It is easy to understand that this would be inefficient, especially for pipelines with complex operations book-ended by other native or complex operations. Such pipelines currently require multiple separate VDMS queries to be invoked.

\begin{figure}[!ht]
    \centering
    \includegraphics[width=0.8\linewidth]{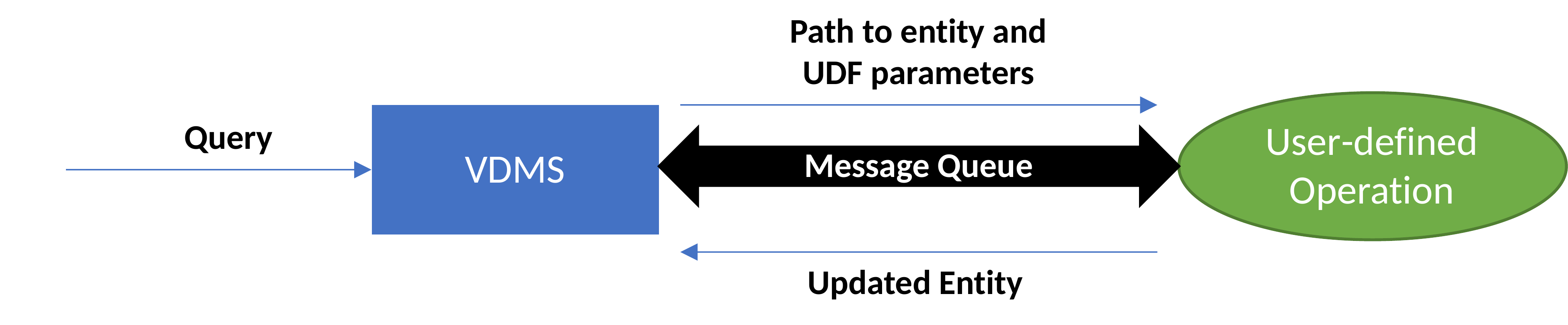}
    \caption{Architecture for User Defined Operations}
    \label{fig:udf_arch}
\end{figure}

User-Defined Operations (UDFs) ensure that users can define their own operations and plug the same into VDMS to execute. As shown in Figure~\ref{fig:udf_arch}, our design utilizes message queues to communicate between VDMS and the user-defined operation running as a separate process. Utilizing the message queue as the inter-process communication mechanism allows asynchronous working of the VDMS server and the UDF process. Also, message queues increase reliability owing to the data persistence at the queues. Moreover, when using message queues, if parallel clients increase the load of the system, the requests can be sent out to the queue without worrying about collision.

\begin{figure}[!ht]
    \centering
    \includegraphics[width=0.5\linewidth]{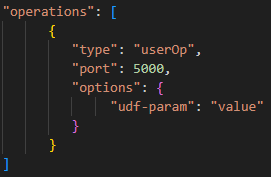}
    \caption{Example User-defined operation}
    \label{fig:udf-example}
\end{figure}

A user can query VDMS the same as in Figure~\ref{fig:query-example}, adding an operation entry for the new user-defined operation as shown in Figure~\ref{fig:udf-example}. The message queue port is an input required from the user to connect VDMS to the same message queue connected to the UDF. The UDF capability also includes an encapsulation to add all UDF parameters inside the $options$ attribute.

\subsection{Remote Operations}
The requirement of remote operations becomes crucial when we consider situations when the VDMS server itself is not competent enough to run compute-intensive operations. Moreover, with increasing data load, one can reduce the execution time by offloading the task to remote servers. Offloading opens the VDMS server for performing other tasks. As discussed earlier, such remoting capability is not available in VDMS. Similar to UDFs, this is a sub-optimal approach contrary to having the support in-built.

\begin{figure}[!ht]
    \centering
    \includegraphics[width=0.8\linewidth]{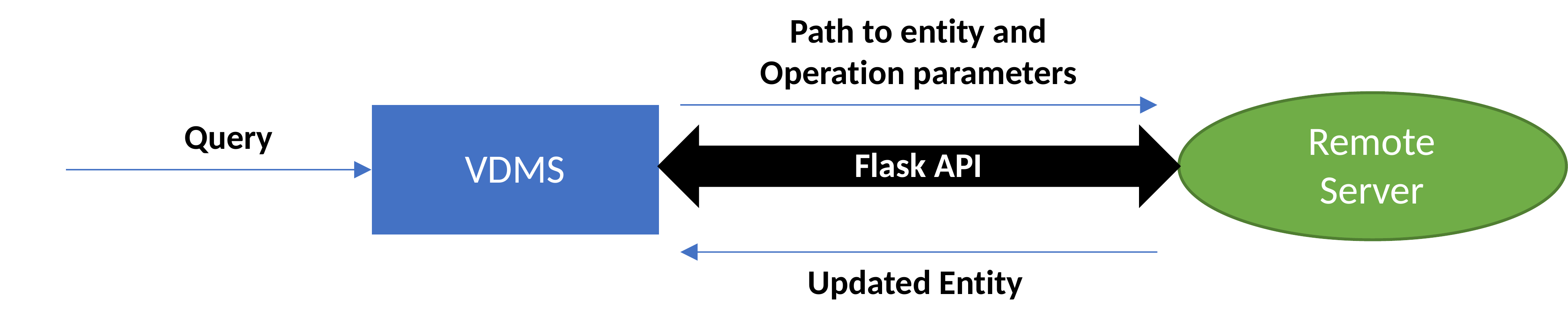}
    \caption{Architecture for Remote Operations}
    \label{fig:remote_arch}
\end{figure}

\begin{figure}[!ht]
    \centering
    \includegraphics[width=0.5\linewidth]{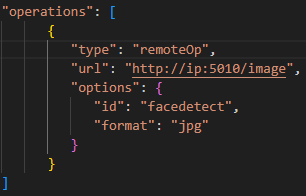}
    \caption{Example Remote operation}
    \label{fig:remote-example}
\end{figure}

We design a similar architecture as UDFs for remote operations where we utilize HTTP APIs (we implemented these using the Flask Python library.) to communicate with the remote server that is running an operation. We transmit the visual data and operation parameters as POST attributes to the remote server and receive the modified visual data files back. A sample remote operation query is shown in Figure~\ref{fig:remote-example}. The URL is the API end-point VDMS will query to execute the operation. These end-points can also be set to execute operations that the user wants to perform, effectively providing a capability to run UDFs on a remote server.


%% file: AsyncModel.tex
\section{Asynchronous Operation Pipeline}\label{asyncmodel}

VDMS follows the classic run-to-completion threading model, where each client query is handled by one thread. This VDMS thread performs operations on one entity at a time, and only when all operations in the query pipeline are executed does it move to the next entity. Moreover, customized user operations might be compute-intensive and hence slow. With the existing synchronous execution model, the VDMS thread sending operations to the remote server would have to idle-wait, delaying the overall execution time. 

A straightforward solution to this stalling problem would be using multiple threads to execute operations on multiple entities. This multi-threaded solution has two major limitations. First, working with extra threads results in extra resource utilization that would put an unnecessary load on the system. Second, such a solution is limited by resource availability and the number of entities in the response. Low-end devices would be unable to support many parallel threads, and thus we would wind back to the same problem of VDMS stalling until the threads are free. Although high-end servers will be able to spawn a larger number of threads, if the number of entities becomes huge, then at some point, we again reach a similar impasse of stalling while waiting for the threads to complete. The other solution could be to utilize batching and multi-threading with remote operations, where each thread handles multiple entities and operations are offloaded to a remote server. But, like the high-end server scenario, the batching approach also effectively delays the evident stalling to a longer time since it does not address the fundamental issue of a thread idle waiting for a remote server to complete the operation on an entity (or batch of entities).

Considering the above challenges, we design an event-driven architecture that considers all operations as events and works asynchronously with only two additional threads.

\begin{figure*}[!ht]
    \centering
    \includegraphics[width=0.8\linewidth]{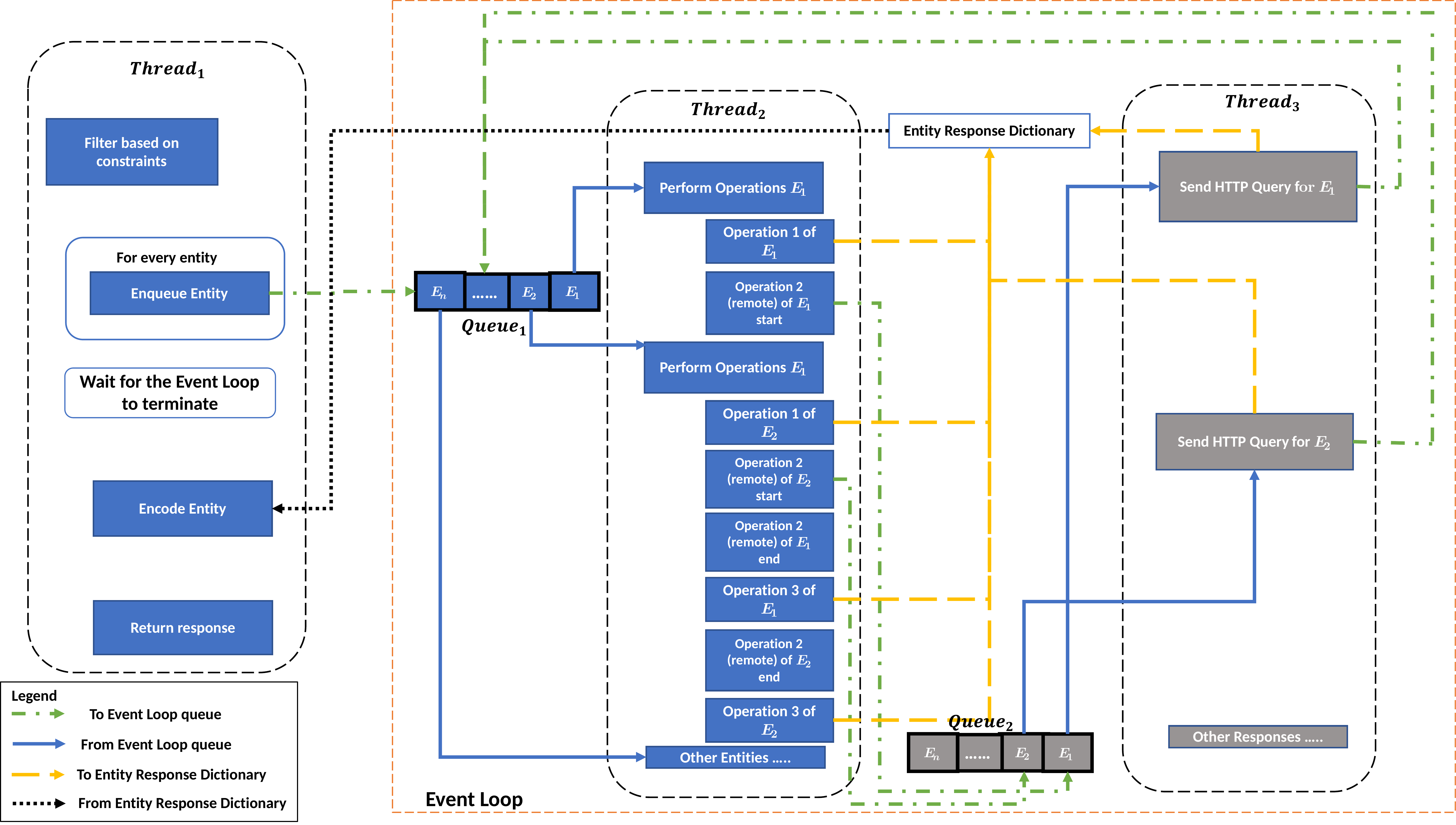}
    \caption{Asynchronous Operation Architecture for \our}
    \label{fig:asyncarch}
\end{figure*}

\subsection{Asynchronous Execution with EventLoops}
We consider every operation as a unique event and modify VDMS to follow an event-driven architecture to execute the operation pipeline. As shown in Figure~\ref{fig:asyncarch}, the architecture is divided into two modules; (a)~the main \our thread, which takes care of communicating with the client, filtering based on constraints, populating the data store, and communicating with the event loop, and (b)~the Event Loop, that catches each event and asynchronously executes them. The event loop maintains two threads to handle two types of events/operations, viz., native ($Thread_2$) and UDF/remote ($Thread_3$). The \our main thread ($Thread_1$) enqueues all entities on $Thread_2$ of the event loop. Following this, the event loop design takes care of how to handle the entities along with their operations asynchronously. 
We next describe both of these sub-modules.

\subsubsection{\textbf{Main \our Thread}}
The main \our thread ($Thread_1$ in Figure~\ref{fig:asyncarch}) takes care of all the tasks except the operation execution, which it offloads to the event loop. $Thread_1$ receives a query from a client and filters out the entities based on the constraints provided by the client. $Thread_1$ maintains the entities as in-memory Visual Compute Library (VCL) objects~\cite{remis2021using}. $Thread_1$ then updates each entity object to add the operations as provided by the client. These entities are then one by one enqueued on the event loop $Queue_1$. It should be noted that not the actual objects but the pointer to these objects are added to the queue. 

Unlike VDMS, \our threads can move on to the following entity instead of waiting for all the operations to complete. Hence, when $Thread_2$ of the event loop starts processing these entities, $Thread_1$ can independently continue enqueueing all the entities. Once all entities are enqueued, $Thread_1$ waits for the event loop to terminate. $Thread_1$ and the event loop use a dictionary as a shared object that includes all updates made to the entities. $Thread_1$ only reads from this dictionary on the event loop's termination and sends back the response to the client.

\subsubsection{\textbf{The Event Loop}}
The event loop works with two queues and two threads to execute the event-driven architecture. $Thread_2$ handles the native VDMS operations, which take less time, while $Thread_3$ handles the UDF/remote operations. Four types of events are considered in the architecture design-- (i)~\textbf{Q1-Enqueue:} An entity is added to $Queue_1$ by $Thread_1$ or $Thread_3$, (ii)~\textbf{R-UDF:} $Thread_2$ encounters a non-native operation in the operation pipeline, (iii)~\textbf{Q2-Enqueue:} An entity is added to $Queue_2$ by $Thread_2$, (iv)~\textbf{R-UDF-Response:} A remote server or the UDF process sends the response back to $Thread_3$. We next describe how each of these events is handled.


\noindent{\textbf{Q1-Enqueue:}} $Thread_2$ keeps track of entities in $Queue_1$ and starts executing operations on the head entity if free. Otherwise, it completes the current execution before dequeuing the next entity. All native operations on these dequeued entities are performed locally within the \our context.

\noindent{\textbf{R-UDF:}} When $Thread_2$ encounters a remote operation or a user-defined operation, it enqueues the entity into $Queue_2$ and releases control on the entity. $Thread_2$ then continues to work on the native operations of the entity currently at the head of $Queue_1$. 

\noindent{\textbf{Q2-Enqueue:}} $Thread_3$ handles the entities that are enqueued in $Queue_2$. It dequeues each entity from the queue and sends it either to the remote server or the UDF process. It then moves on to the next entity in the queue. 

\noindent{\textbf{R-UDF-Response:}} A callback method on $Thread_3$ listens for a response on the entities sent to a remote server or the UDF process. Once an entity's response arrives, this callback method updates the $Entity\,Response\,Dictionary$ ($ERD$) and enqueues the entity back to $Queue_1$ to be processed by $Thread_2$. It should be noted that both $Thread_2$ and $Thread_3$ work on different entities at a time and hence do not cause any conflict. Similarly, the $ERD$ is also updated by the threads for different entities at any time. There is a possibility of conflict between $Thread_1$ and $Thread_3$ while enqueueing $Queue_1$, which is handled by a standard lock mechanism.

\begin{figure*}[!ht]
    \centering
    \includegraphics[width=0.8\linewidth]{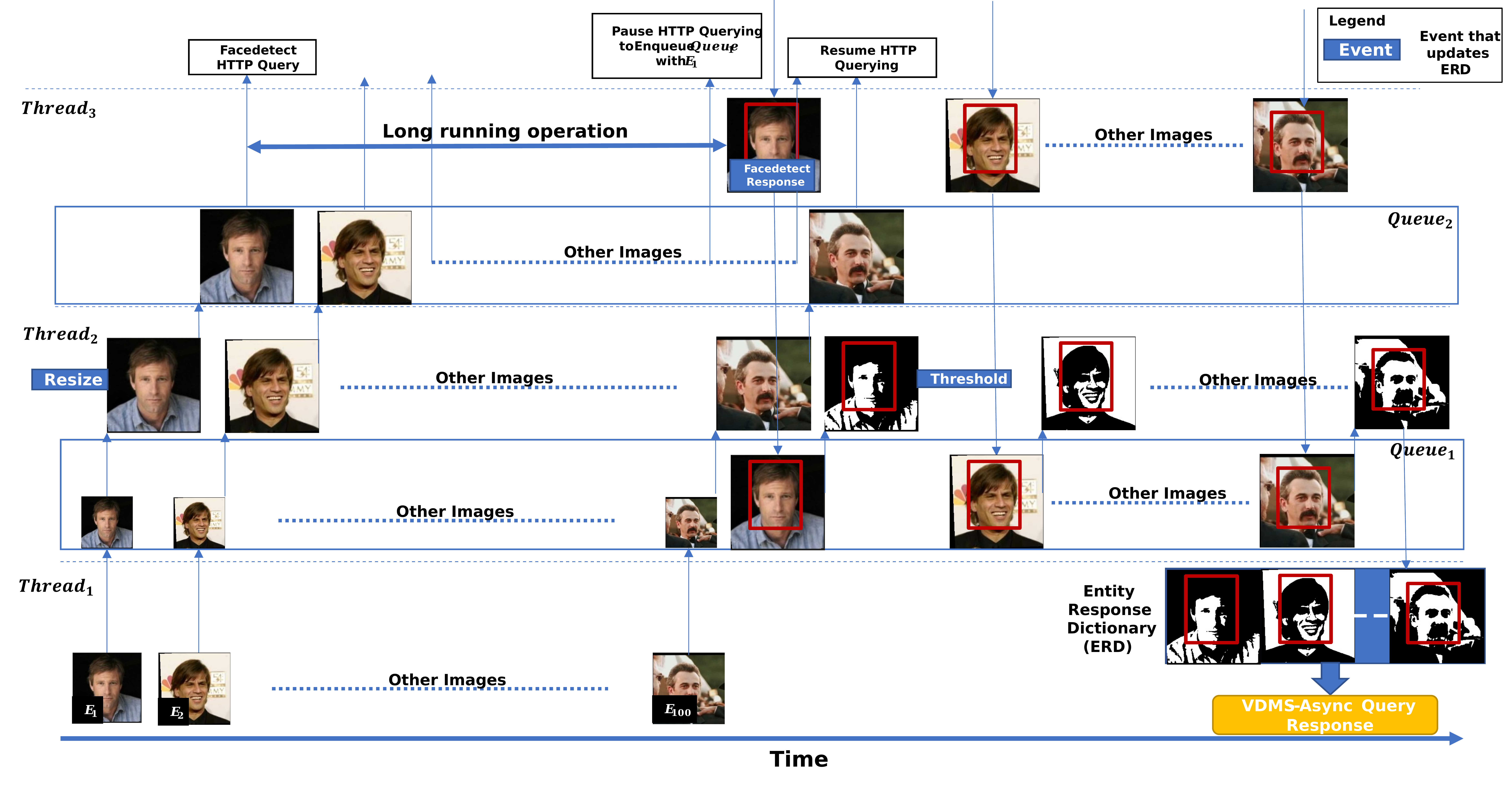}
    \caption{An illustration of how use-case in Section~\ref{usecase} is by \our. Multiple updates of $Entity\,Response\,Dictionary$ ($ERD$) not shown for simplicity. Images are used from the Labeled Faces in the Wild~\cite{huang2014labeled} dataset. (Not to scale)}
    \label{fig:usecase_demo}
\end{figure*}

\subsection{Example Use-Case}\label{usecase}
Consider a user wants to query the \our data store for images with a set of constraints-- ($category = celebrity$ and $21 \leq age \leq 40$) and operations-- Resize image (Native) $\rightarrow$ Detect and mark face (Remote) $\rightarrow$ Threshold image (Native)

\begin{figure}[!ht]
    \centering
    \includegraphics[width=0.4\linewidth]{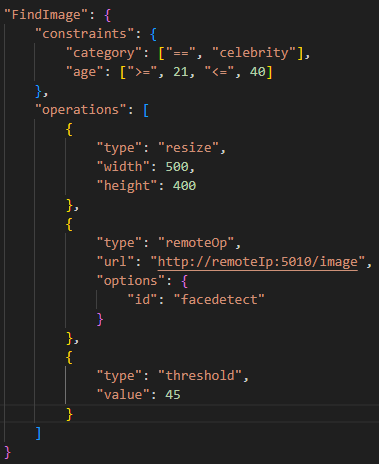}
    \caption{\our query for the use-case in Section~\ref{usecase}}
    \label{fig:usecase}
\end{figure}

The corresponding \our query is as shown in Figure~\ref{fig:usecase}. We illustrate the working of \our to perform the query in Figure~\ref{fig:usecase_demo}. The details are described next. 

On receiving this query, $Thread_1$ filters out the images (say $100$ images based on the constraints) and adds the three operations (resize, facedetect, and threshold) to all the VCL image objects. $Thread_1$ then starts enqueueing pointers to the image objects to $Queue_1$ and continues the task until all $100$ images have been enqueued. 

In the meantime, $Thread_2$ starts operating on the first image, resizes the image as per the operation parameters (400, 500), and updates the $ERD$. Next, it encounters a remote operation and enqueues the pointer to the image object on $Queue_2$. Thus, $Thread_2$ releases the first image and continues with the second to perform the resize operation. After resizing the second image, it updates the $ERD$, encounters the remote operation for the second image and enqueues the pointer to the image on $Queue_2$. It then moves on to the third image and so on.

Meanwhile, $Thread_3$ encounters an update to $Queue_2$ and creates an HTTP Post request to query the remote server for the first image. It makes the query, releases the first image and moves on to the second image. As $Thread_2$ keeps populating $Queue_2$, $Thread_3$ keeps making the HTTP requests. Whenever there is a response from the remote server for an image, note that this need not be in order as some images might take more time for operation execution, $Thread_3$ pauses the HTTP query context and executes a callback that updates the image for which the response has arrived. It also updates the $ERD$ with the updated image. $Thread_3$ then enqueues this image onto $Queue_1$ and switches back to the HTTP querying task. When $Thread_2$ encounters the image for which remote operation has been completed, it then executes the threshold operation on it and updates the $ERD$. We update the $ERD$ after every operation to ensure we have the latest updates to the entity if any failure occurs. After the third operation is executed, the image is released by all threads. Once both $Queue_2$ and $Queue_3$ are empty, and both $Thread_2$ and $Thread_3$ are free, the event loop is terminated.

On the event loop's termination, $Thread_1$ starts retrieving the images from the updated $ERD$ and processes the images to finally send a response to the user.

\subsection{Scaling-out with Remoting}
The availability of remote operations gives \our the advantage of offloading the operation execution to a remote server. Furthermore, the event-driven architecture makes offloading to multiple servers at the same time feasible. Consider a situation where a large number of clients are querying \our at the same time. Although we can offload a task to one server asynchronously, the remote server itself will become a bottleneck owing to its own resource constraints. Especially when the query itself is compute-intensive. If, instead, we have an ecosystem of $\kappa$ such remote servers, then theoretically, the workload on the remote server will be reduced by $\kappa$. This workload reduction will lead to $~\kappa$ times reduction in the query execution time too.

The capability to expand to $\kappa$ remote servers to offload operations asynchronously helps to scale out \our to serve a much larger number of entities and parallel clients at the same time.

%% file: Evaluation.tex
\section{Evaluation}\label{evaluation}
In this section, we evaluate different aspects of \our covering query execution time, resource utilization, and how \our fares with respect to the existing systems. We evaluate \our with both image and video datasets. Furthermore, we also analyze the improvements we gain with \our when we scale out by using multiple remote servers.

\subsection{Image Dataset}
We next describe the Labeled Faces in the Wild~\cite{huang2014labeled} dataset that we use to evaluate \our with images and the queries we use to perform the benchmarking.


\subsubsection{\textbf{Dataset Description}}
The Labeled Faces in the Wild (LFW) dataset~\cite{huang2014labeled} is a public benchmarking dataset with 13000 images of human faces. We add images into \our using the $AddImage$ query and two metadata entries, viz., $name$, provided as part of the dataset and $category$ as $"lfw"$. 

\subsubsection{\textbf{Benchmarking Queries}}
We design a set of nine benchmarking queries that are described below. Each of these queries first filters images based on user constraints and then performs the specific operation described in the query description. Unless specified, all these queries are performed as a UDF on a remote server, and the parameters provided by the users are defined in the $options$ attribute of the operation entry.

\begin{itemize}
    \item \textbf{Crop (IQ1)} - Crop the image to a size of ($height, width$), starting from the coordinate ($x,y$).
    \item \textbf{Grayscale (IQ2)} - Change the image colour to grayscale.
    \item \textbf{Blur (IQ3)} - Blur the image using Gaussian Blur. The user provides the $height$ and $width$ of the Gaussian kernel that will be used to blur the image, both of which should be positive and odd. The user can also provide the standard deviation, $SigmaX$ and $SigmaY$ (default value is 0 if not specified).
    \item \textbf{Box (IQ4)} - Detect a face in the image and create a box around it.
    \item \textbf{Mask (IQ5)} - Create a circular mask at the centre of the face in the image of radius $r$ provided by the user.
    \item \textbf{Upsample (IQ6)} - Increase image resolution by $X$ times along x-axis and $Y$ times along y-axis. $X$ and $Y$ are specified by the user.
    \item \textbf{Downsample (IQ7)} - Decrease image resolution by $X$ times along x-axis and $Y$ times along y-axis. $X$ and $Y$ are specified by the user.
    \item \textbf{Caption (IQ8)} - Add text to the image at user specified coordinates. The user provides both text and coordinates.
    \item \textbf{Manipulation (IQ9)} - Identify a face in the image, create a circular mask around it and black out all other parts of the image such that only the face is visible.
\end{itemize}


\subsection{Video Dataset}
We next describe the Kinetics400 Human Action Video Dataset~\cite{kay2017kinetics} that we use to evaluate \our with videos and the queries we use to perform the benchmarking.

\subsubsection{\textbf{Dataset Description}}
The Kinetics400 Human Action Video Dataset~\cite{kay2017kinetics} is designed to detect different types of human actions in a video. The Kinetics400 version has 306,245 videos, each lasting around 10 seconds. We add videos into \our using the $AddVideo$ query. We add two metadata entries to each video; (a)~$category$ as $"activity"$, and (b)~$activity$, which specifies the activity being performed in the video. 

\subsubsection{\textbf{Benchmarking Queries}}
We design a set of nine benchmarking queries that are described below. Each of these queries first filters videos based on user constraints and then performs the specific operation described in the query description. Unless specified, all these queries are performed as a UDF on a remote server, and the parameters provided by the users are defined in the $options$ attribute of the operation entry.

\begin{itemize}
    \item \textbf{Select (VQ1)} - Select an interval in the video between $t_1$ and $t_2$ provided by the user. Then crop the frames in the interval into rectangles of size ($height$, $width$), starting at coordinates ($x,y$), both of which are provided by the user.
    \item \textbf{Grayscale (VQ2)} - Convert the whole video into grayscale.
    \item \textbf{Blur (VQ3)} - Blur the whole video using Gaussian Blur. The user provides the $height$ and $width$ of the Gaussian kernel that will be used to blur the video frames, both of which should be positive and odd. The user can also provide the standard deviation, $SigmaX$ and $SigmaY$ (default value is 0 if not specified).
    \item \textbf{Box (VQ4)} - Detect faces in all the video frames and mark them with a box. The resulting video should have these boxes on all faces in all frames.
    \item \textbf{Mask (VQ5)} - Detect faces in all the video frames and create a circular mask at the centre of the faces in the frames of radius $r$ provided by the user.
    \item \textbf{Upsample (VQ6)} - Increase video resolution by $X$ times along x-axis and $Y$ times along y-axis. $X$ and $Y$ are specified by the user.
    \item \textbf{Downsample (VQ7)} - Decrease video resolution by $X$ times along x-axis and $Y$ times along y-axis. $X$ and $Y$ are specified by the user.
    \item \textbf{ActivityRecognition (VQ8)} - Detect an activity in a video frame and add a text in the frame specifying the activity. The resulting video should have activity text in all frames.
    \item \textbf{Manipulation (VQ9)} - Identify a face in video frames, create a circular mask around it and black out all other parts of the frame such that only the face is visible.
\end{itemize}

\subsection{Competing Systems}
As discussed in Section~\ref{background}, no other existing systems provide the visual data management capabilities provided by VDMS for both images as well as videos. Instead, relational database systems like MySQL and PostgreSQL have support to work with videos and images. However, the support is very limited with MySQL when considering user-defined operations. Hence, we utilize the PostgreSQL relational database system to compare the effectiveness of \our along with VDMS.

\subsubsection{\textbf{PostgreSQL}~\cite{momjian2001postgresql}}: PostgreSQL is an open-source object-relational database system frequently utilized in the literature. Unlike MySQL, PostgreSQL has more flexibility for defining user-defined operations as PostgreSQL functions. Furthermore, PostgreSQL supports concurrent writing without locks and implements transaction isolation, leading to faster parallel UDF execution. Existing works have also shown that PostgreSQL can be used for working with visual data operations~\cite{he2020method}. We implement the UDFs required for the benchmarking queries as internal PostgreSQL functions.

\subsubsection{\textbf{VDMS}~\cite{remis2021using}}: The existing version of VDMS acts as another competing system. As VDMS does not support user-defined operations, we ensure a fair comparison by including the UDF and remote operations capability in the existing version of VDMS for the experiments. Unlike \our, VDMS follows a synchronous operation execution approach.

\subsubsection{\textbf{Scanner}~\cite{poms2018scanner}}:\label{scanner_compete} Scanner is a distributed visual data management system that is designed to suit video processing applications at scale. Scanner takes every frame as a separate entity, and operations are performed on a frame-by-frame basis. Scanner creates a computation graph model to design the workflow that includes extracting frames from the video, performing operations on the frames, and generating the resulting video. Optimization is achieved by employing scheduling, parallelization, and frame compression to H.264 byte streams~\cite{marpe2006h} to speed up the video processing tasks. Scanner does not support cropping (VQ1), activity recognition (VQ8), and manipulation (VQ9). Also, video resizing (VQ6-7) is only supported with static height and width. Hence, we design our own operations and incorporate them into Scanner for the tests. As Scanner does not support images, we only provide results for video benchmarking.

\subsection{Experiment Setup}
We run all competing systems, including \our, on an Ubuntu 18.04 dual socket server. The server runs with Intel\textregistered Xeon\textregistered Gold 6252 CPU with 96 cores @ 2.10GHz and has 376 GB of DDR4 DRAM. The remote server we use for remote operations runs a dual-socket Ubuntu 22.04 server with  Intel\textregistered Xeon\textregistered Gold 6140 CPU with 35 cores  @ 2.3 GHz and has 384 GB of DDR4 DRAM. A client making a query runs on a third machine running a dual-socket Ubuntu 22.04 server with  Intel\textregistered Xeon\textregistered Platinum 8180 CPU with 28 cores @ 2.5 GHz and has 188 GB of DDR4 DRAM

We utilize three metrics, viz. query execution time, entities processed per time unit and CPU utilization, to evaluate \our and the competing systems. We perform separate experiments for images and videos, which means no single query expects both images and videos in the response. We execute every query for $15$ independent runs and average over the metrics to report the results. We also analyze the results when the systems have to support multiple clients concurrently. In our experiments, we vary the number of concurrent clients from 2 to 128 by doubling the number of clients every time. As our analysis is more dependent on the operation execution capability of the systems rather than the size of the dataset only, for all our experiments, we use a subset of the datasets with 5000 images and 500 videos selected at random. We perform three categories of experiments for both datasets;

\begin{itemize}
    \item[\textbf{C1}] One client executes one of the nine queries without constraints. The response is all the entities (5000 images or 500 videos) modified by the operation specified in the query.    
    \item[\textbf{C2}] One client executes a custom query with no constraints but multiple operations. The response is all the entities modified by the operation pipeline. The following pipeline is used; 
    \begin{itemize}
        \item \textbf{Image}: Resize $\rightarrow$ Box $\rightarrow$ Manipulation $\rightarrow$ Rotate. Here, Resize and Rotate are native operations performed locally.
        \item \textbf{Video}: Activity Recognition $\rightarrow$ Resize $\rightarrow$ Select $\rightarrow$ Manipulation. Here, Resize is an operation performed locally.
    \end{itemize}
    \item[\textbf{C3}] Multiple clients execute the same custom query as C2 at the same time. It should be noted that the total number of entities that match the query and require processing increases proportionally with multiple clients. For example, the systems must process $40k$ images and $4k$ videos with eight clients.
\end{itemize}

\subsection{Evaluation of \our: Image Dataset}
We evaluate the systems using the image benchmarking queries over the LFW dataset for each of the three categories of experiments (C1 - C3).

\begin{figure}[!ht]
    \centering
    \includegraphics[width=0.7\linewidth]{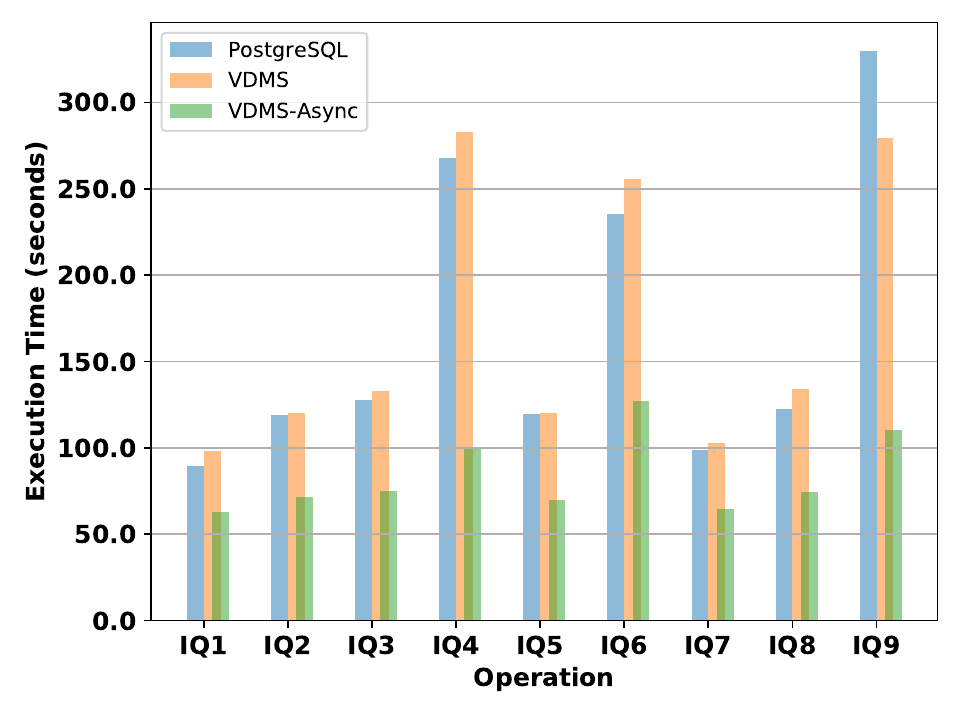}
    \caption{[Image-C1]: Query Duration}
    \label{fig:sqme}
\end{figure}

\begin{figure}[!ht]
    \centering
    \includegraphics[width=0.7\linewidth]{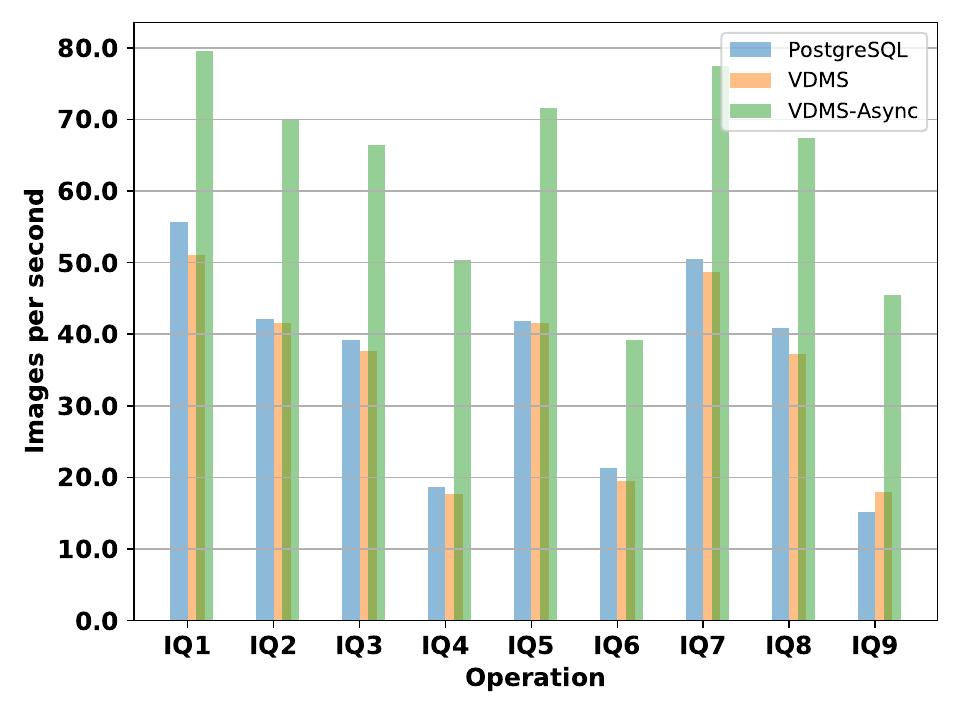}
    \caption{[Image-C1]: Throughput}
    \label{fig:ips_sqme}
\end{figure}

\begin{figure}[!ht]
    \centering
    \includegraphics[width=0.7\linewidth]{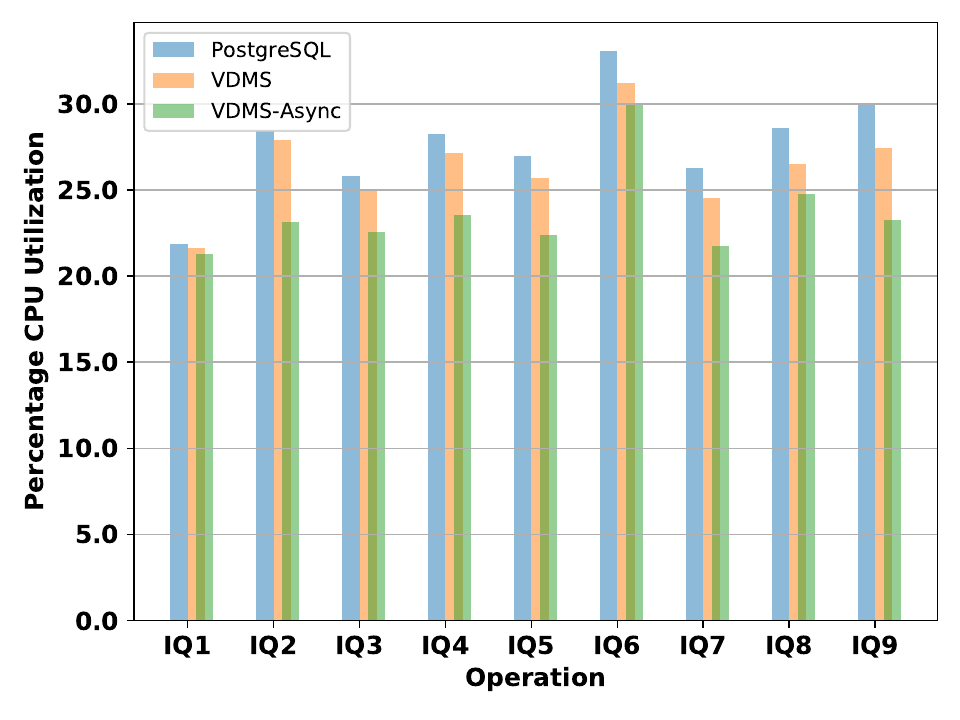}
    \caption{[Image-C1]: Average Resource Usage }
    \label{fig:cpu_sqme}
\end{figure}

\subsubsection{\textbf{Image-C1:} }In the first set of experiments, we query the system with each of the nine benchmarking queries one at a time.

As is evident from Figure~\ref{fig:sqme}, VDMS takes the maximum time to execute the query owing to the absence of any parallelization. Although PostgreSQL gains compared to VDMS due to its ability to execute functions in parallel, the gain is not too high as there are limitations to the number of parallel transactions that can be generated. \our, on the other hand, gains considerably compared to the other two systems because of the event-driven architecture that minimizes any stalling that could occur. 

Using an event loop where we switch to another image instead of waiting ensures that \our can process a higher number of images for all the benchmarking queries (Figure~\ref{fig:ips_sqme}).

With equivalent work being done by all systems, the CPU utilization is nearly similar for all three (Figure~\ref{fig:cpu_sqme}).


\begin{figure}[!ht]
    \centering
    \includegraphics[width=0.7\linewidth]{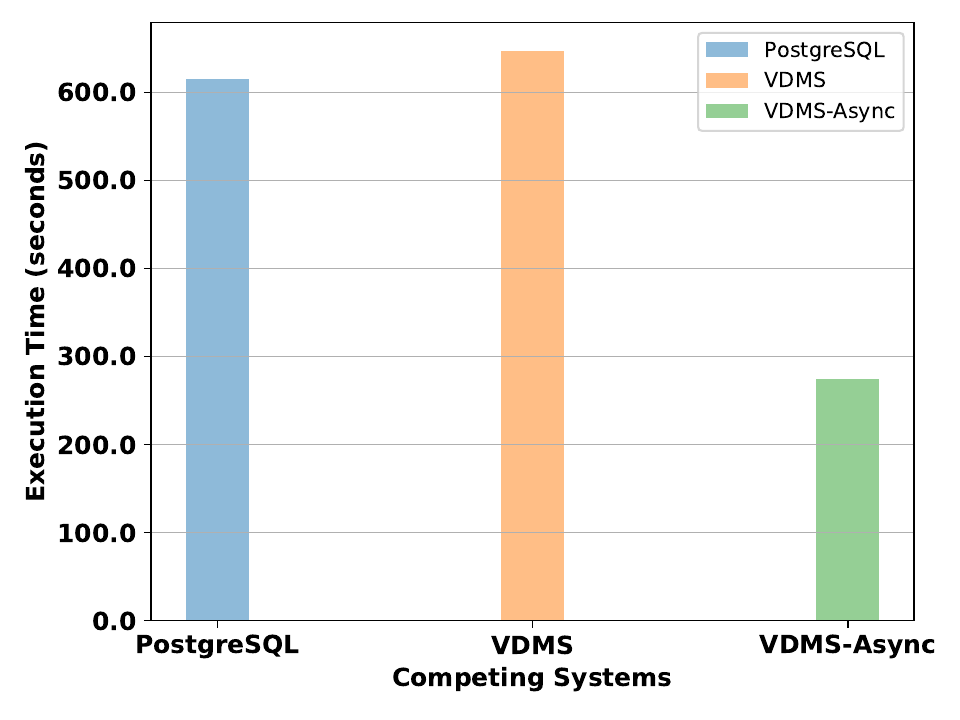}
    \caption{[Image-C2]: Query Duration}
    \label{fig:pme}
\end{figure}

\begin{figure}[!ht]
    \centering
    \includegraphics[width=0.7\linewidth]{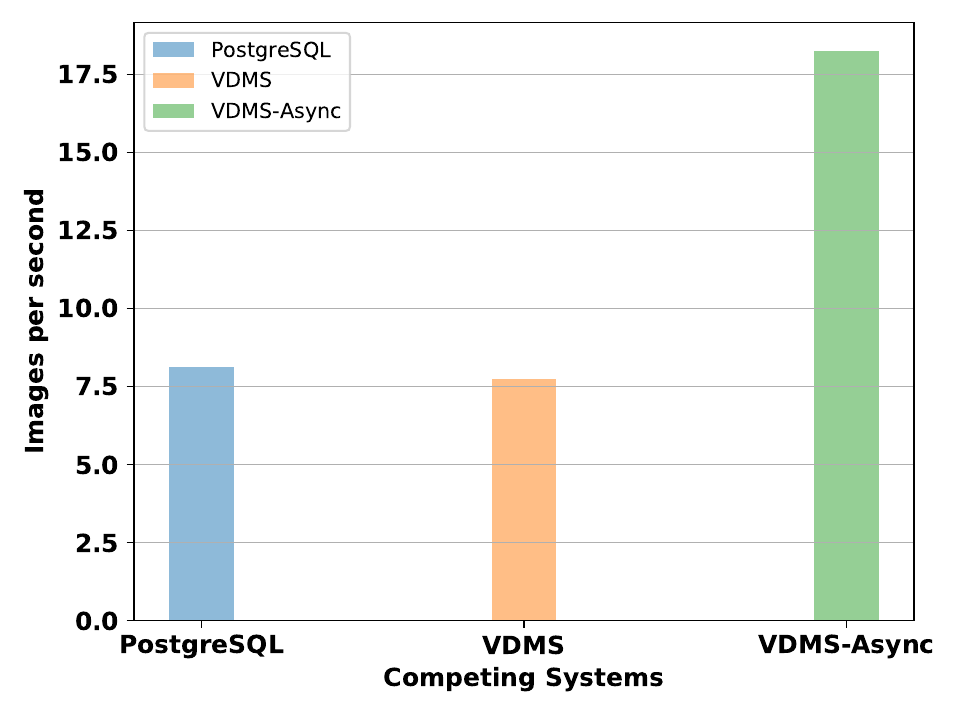}
    \caption{[Image-C2]: Throughput}
    \label{fig:ips_pme}
\end{figure}

\begin{figure}[!ht]
    \centering
    \includegraphics[width=0.7\linewidth]{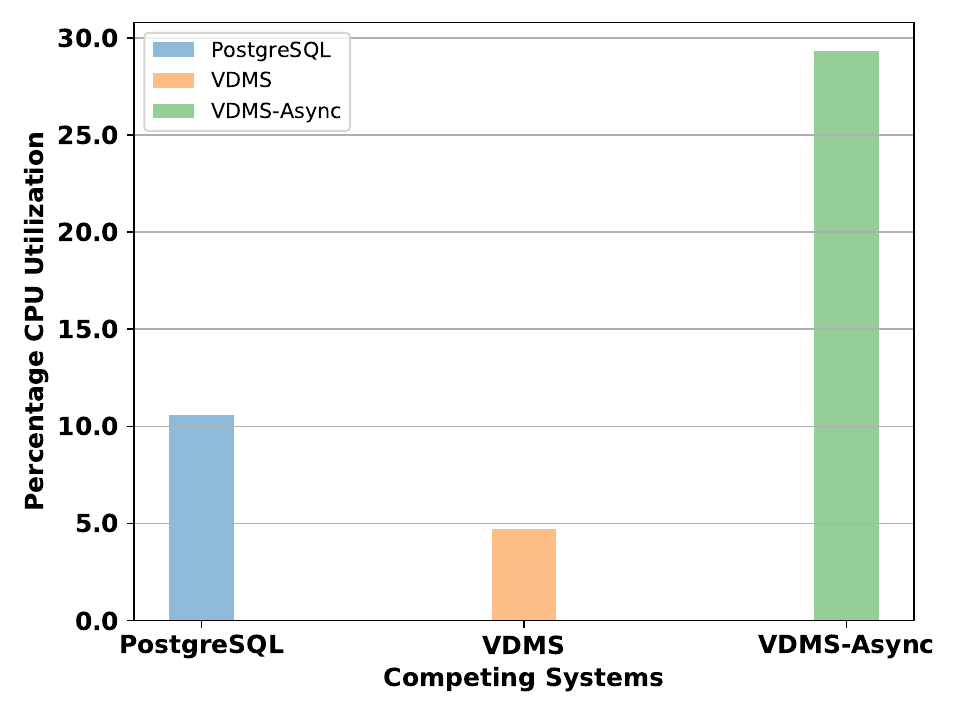}
    \caption{[Image-C2]: Average Resource Usage}
    \label{fig:cpu_pme}
\end{figure}

\subsubsection{\textbf{Image-C2:} }In the next set of experiments, we query the system with an operation pipeline, Resize $\rightarrow$ Box $\rightarrow$ Manipulation $\rightarrow$ Rotate.

Gains of \our with respect to the other two systems are much higher compared to C1 for query duration (Figure~\ref{fig:pme}) and throughput (Figure~\ref{fig:ips_pme}) because of the higher workload leading to longer stalling and idle waits for VDMS and PostgreSQL. \our has an overall gain of $~2.5$X compared to both VDMS and PostgreSQL in terms of query duration. \our can also process $>2X$ images per second than the other competing systems. Such gains with \our are linked to the fact that when executing an operation pipeline, the HTTP requests to the remote server and the native operations execute simultaneously on separate threads on different images. So, neither the VDMS server nor the remote server has to idle-wait. On the contrary, VDMS can only work with one image at a time and idle waits when the operation is being performed at the remote server. Similarly, with VDMS, the remote server has to idle-wait when VDMS is executing the native operations. Due to these idle wait periods, VDMS spends more time executing the entire operation pipeline per entity.

Although PostgreSQL can spawn multiple processes under load to enable parallelism, providing a slight edge over VDMS, the number of available CPUs limits this vector of parallelism. More importantly, these processes still need to wait for a response from the remote server before proceeding to the subsequent entities matching the query constraints. Similarly, the remote server has to idle-wait for the next set of images when PostgreSQL performs native operations on the current set of images.

We see a change in the trend for CPU Utilization (Figure~\ref{fig:cpu_pme}) owing to the idle-wait times in VDMS and PostgreSQL. Since the servers running VDMS and PostgreSQL are waiting to receive the response from the remote server, the CPU utilization is close to zero. However, for \our, there is negligible idle waiting involved, and all the threads are active throughout the processing pipeline unless their task is complete. This leads to a higher CPU utilization for \our compared to the competing systems.

\begin{figure}[!ht]
    \centering
    \includegraphics[width=0.7\linewidth]{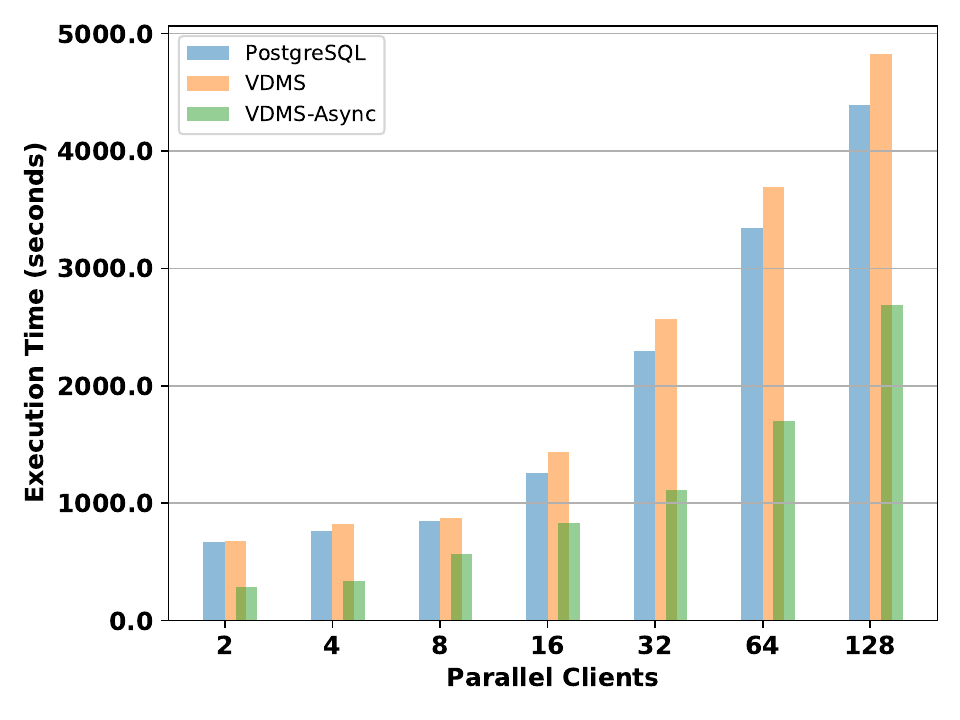}
    \caption{[Image-C3]: Query Duration}
    \label{fig:mcp}
\end{figure}

\begin{figure}[!ht]
    \centering
    \includegraphics[width=0.7\linewidth]{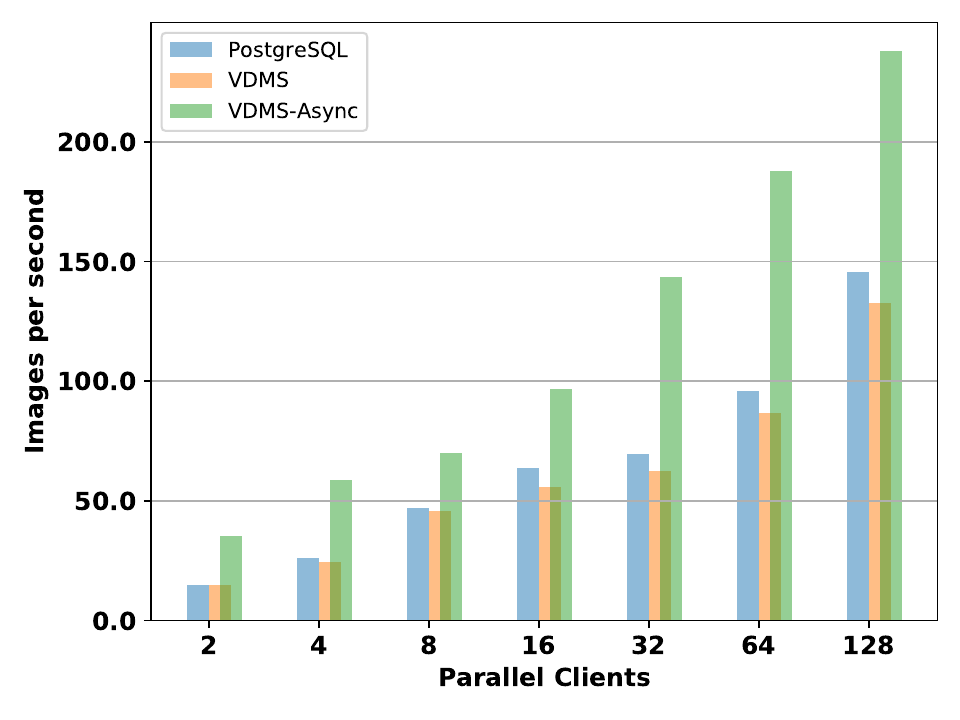}
    \caption{[Image-C3]: Throughput}
    \label{fig:ips_mcp}
\end{figure}

\begin{figure}[!ht]
    \centering
    \includegraphics[width=0.7\linewidth]{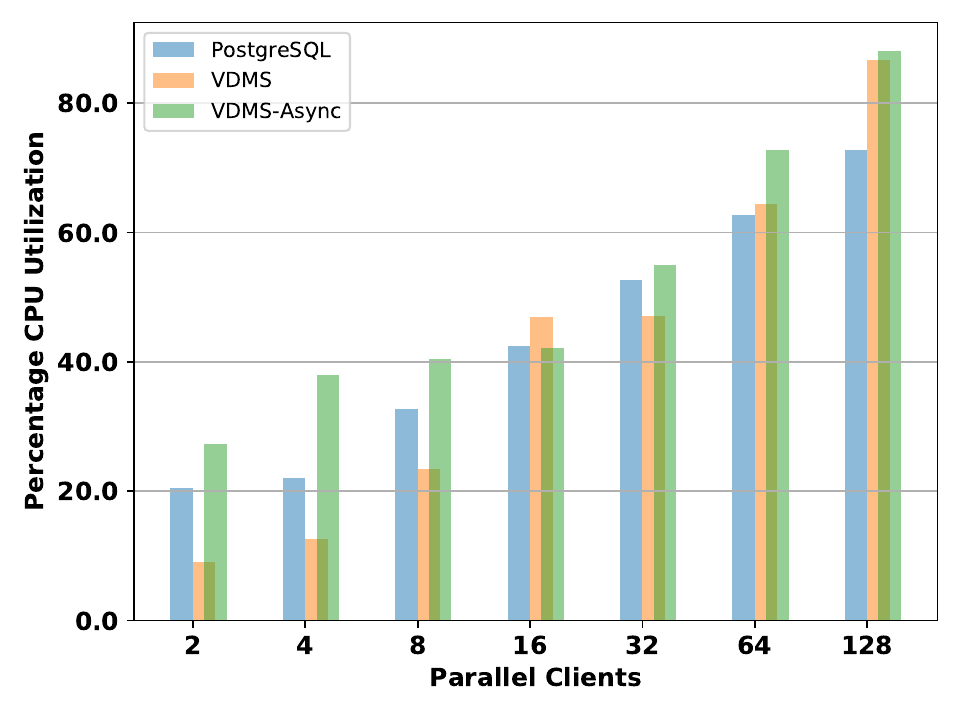}
    \caption{[Image-C3]: Average Resource Usage}
    \label{fig:cpu_mcp}
\end{figure}

\subsubsection{\textbf{Image-C3:} }We next query the operation pipeline used in C2, but now multiple clients query the system concurrently. We increase the number of concurrent clients from 2 to 128 by doubling every run.

With multiple clients, \our achieves much lower execution time (~$2$X as shown in Figure~\ref{fig:mcp}) and doubles the number of images processed per second (Figure~\ref{fig:ips_mcp}). We see similar trends as C2 for CPU utilization (Figure~\ref{fig:cpu_mcp}). However, the difference in CPU utilization between the three systems starts diminishing with increasing clients as PostgreSQL and VDMS get much lower idle time because of the continuous processing of requests from the higher number of clients.

\subsection{Evaluation of \our: Video Dataset}

We evaluate the systems using the video benchmarking queries over the Kinetics dataset for each of the three categories of experiments (C1 - C3).

\begin{figure}[!ht]
    \centering
    \includegraphics[width=0.7\linewidth]{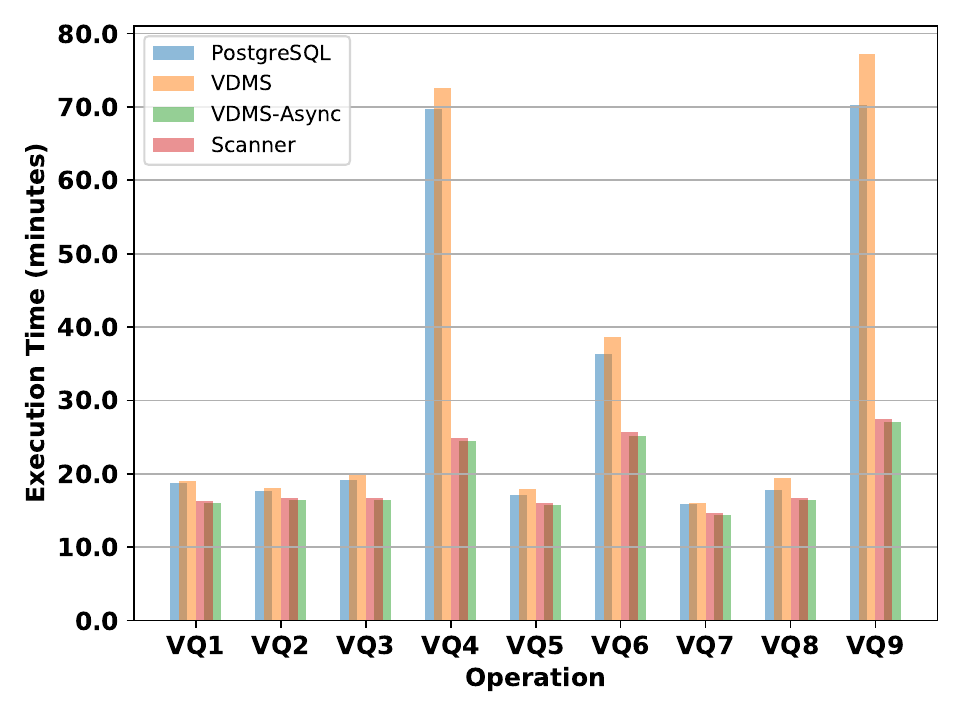}
    \caption{[Video-C1]: Query Duration}
    \label{fig:vsqme}
\end{figure}

\begin{figure}[!ht]
    \centering
    \includegraphics[width=0.7\linewidth]{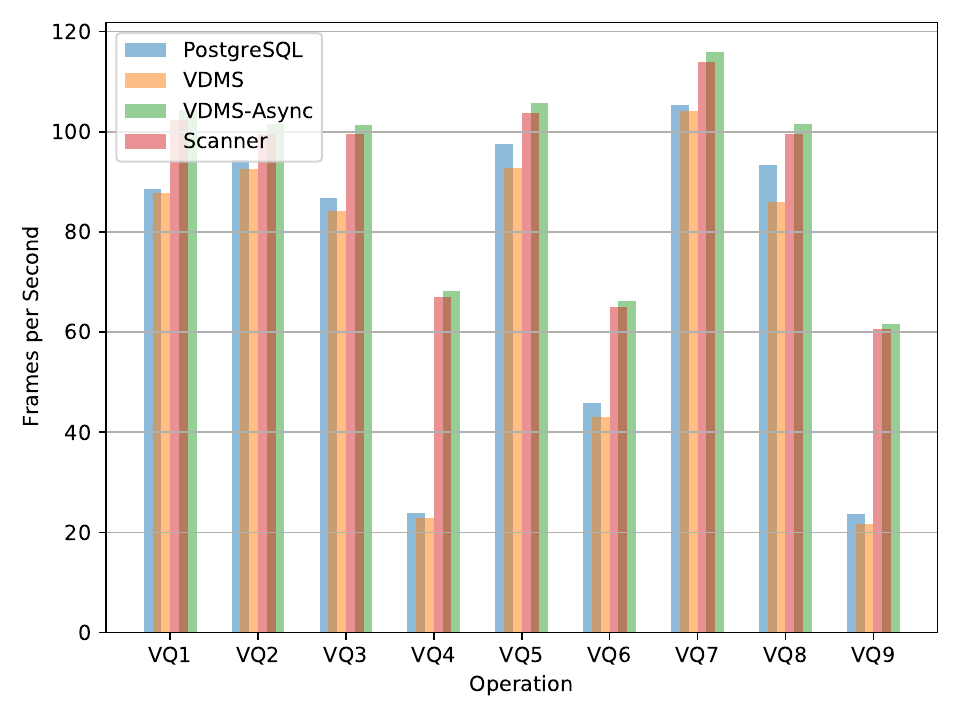}
    \caption{[Video-C1]: Throughput}
    \label{fig:vps_sqme}
\end{figure}

\begin{figure}[!ht]
    \centering
    \includegraphics[width=0.7\linewidth]{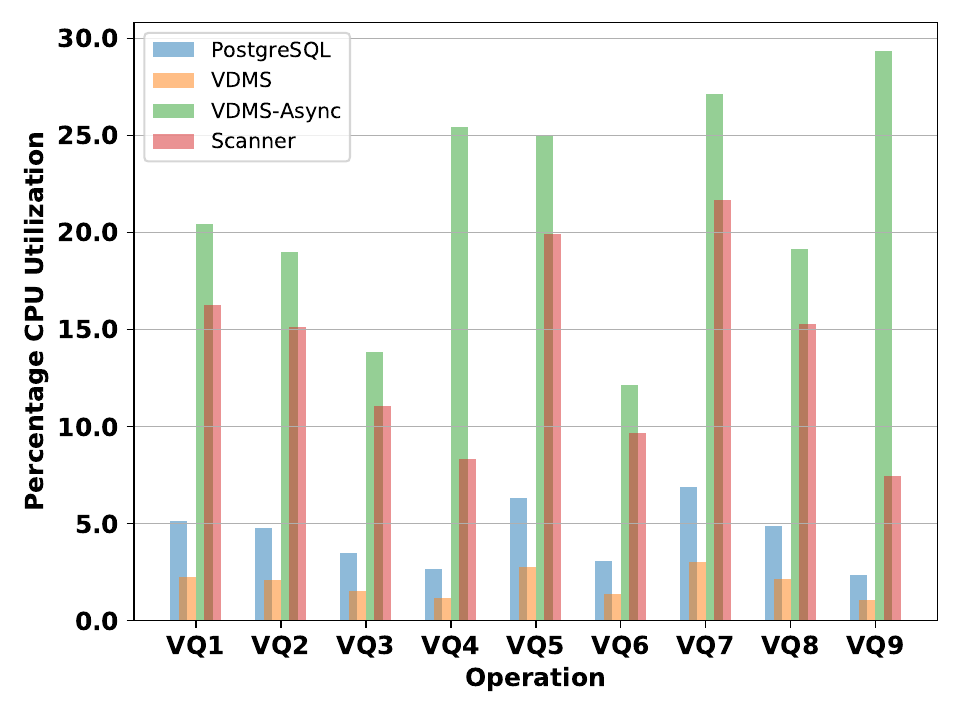}
    \caption{[Video-C1]: Average Resource Usage}
    \label{fig:vcpu_sqme}
\end{figure}

\subsubsection{\textbf{Video-C1:} }In the first set of experiments, we query the system with each of the nine benchmarking queries one at a time.

We see similar trends with the video dataset (Figure~\ref{fig:vsqme}) as we saw for the image dataset\footnote{It should be noted that we have reported the execution time here in minutes, and hence the relative difference looks less compared to images.}. The gain is much higher for compute-intensive tasks such as face detection (VQ4) and video manipulation (VQ9), for which the competing systems PostgreSQL and VDMS take $3X$ the time than \our to complete the query. Scanner performs similarly to \our owing to the speed-up provided by the computation graph (\S~\ref{scanner_compete}).

The relative gain for the number of frames processed per second is higher than we have seen for images (Figure~\ref{fig:vps_sqme}). For instance, the number of images processed for IQ9 is $~45$, while for VQ9, the number of frames processed per second is $~62$. A higher gain is observed with frames as the operations receive the frames packaged as a video in a single call, while for images, there has to be a separate call every time, resulting in greater communication overheads. Overall, \our fares better than the PostgreSQL and VDMS with higher gains for compute-intensive queries.

Again a more compute-intensive task than operating over images leads to more compute requirements for videos (Figure~\ref{fig:vcpu_sqme}). Longer idle times for PostgreSQL and VDMS result in negligible CPU utilization during those periods leading to lower average CPU Utilization. On the contrary, Scanner that is always active owing to working frame by frame has higher CPU utilization. Compared to Scanner, \our too is always active, but with more threads and hence shows a higher resource utilization.

\begin{figure}[!ht]
    \centering
    \includegraphics[width=0.7\linewidth]{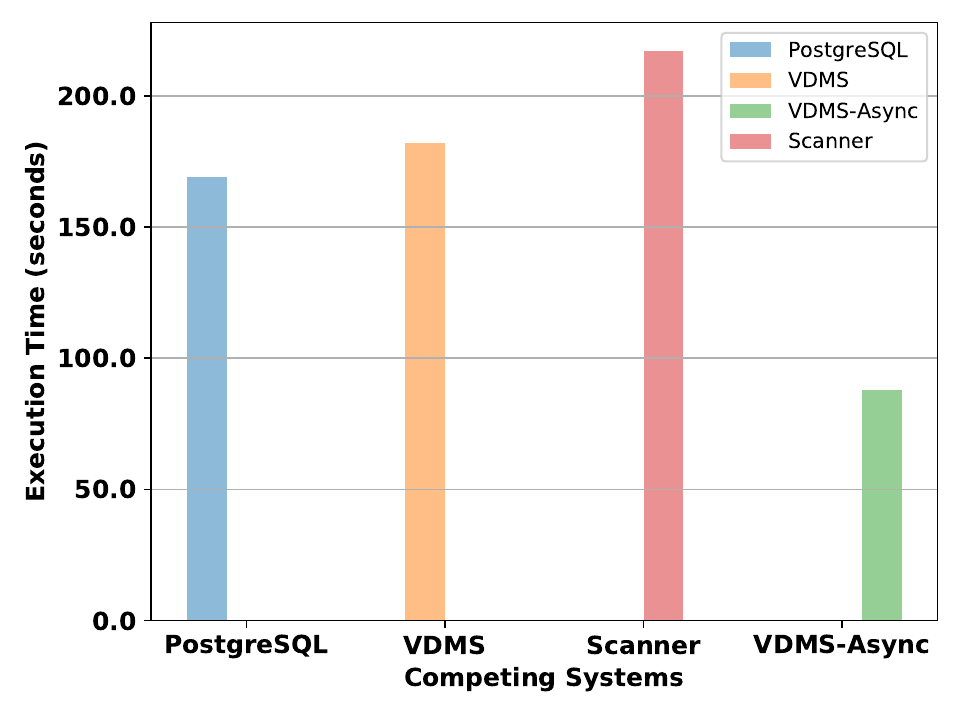}
    \caption{[Video-C2]: Query Duration}
    \label{fig:vpme}
\end{figure}

\begin{figure}[!ht]
    \centering
    \includegraphics[width=0.7\linewidth]{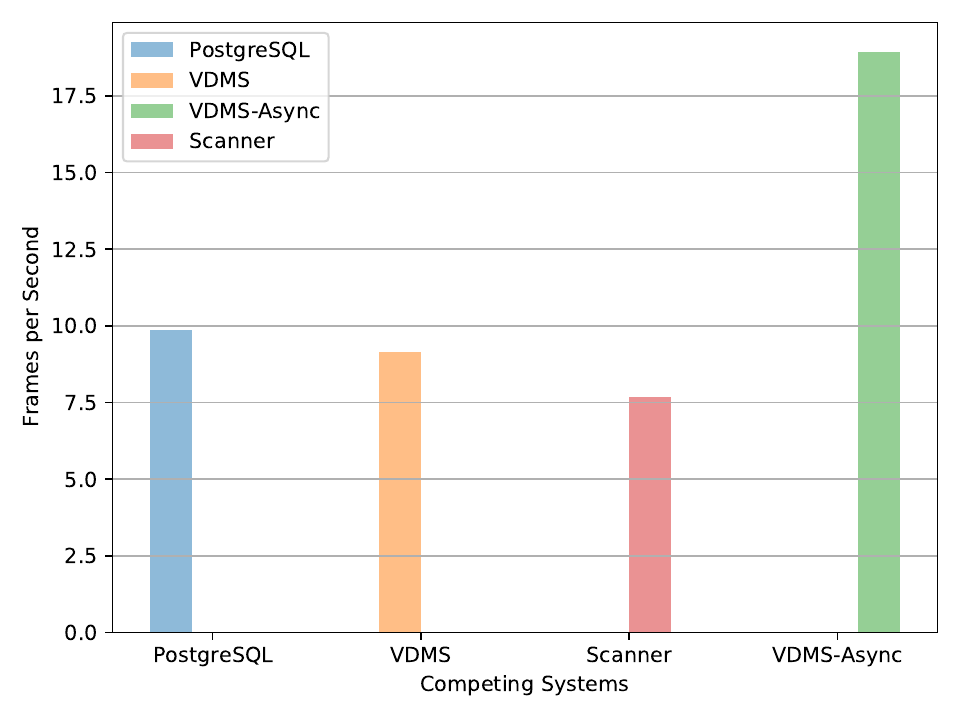}
    \caption{[Video-C2]: Throughput}
    \label{fig:vps_pme}
\end{figure}

\begin{figure}[!ht]
    \centering
    \includegraphics[width=0.7\linewidth]{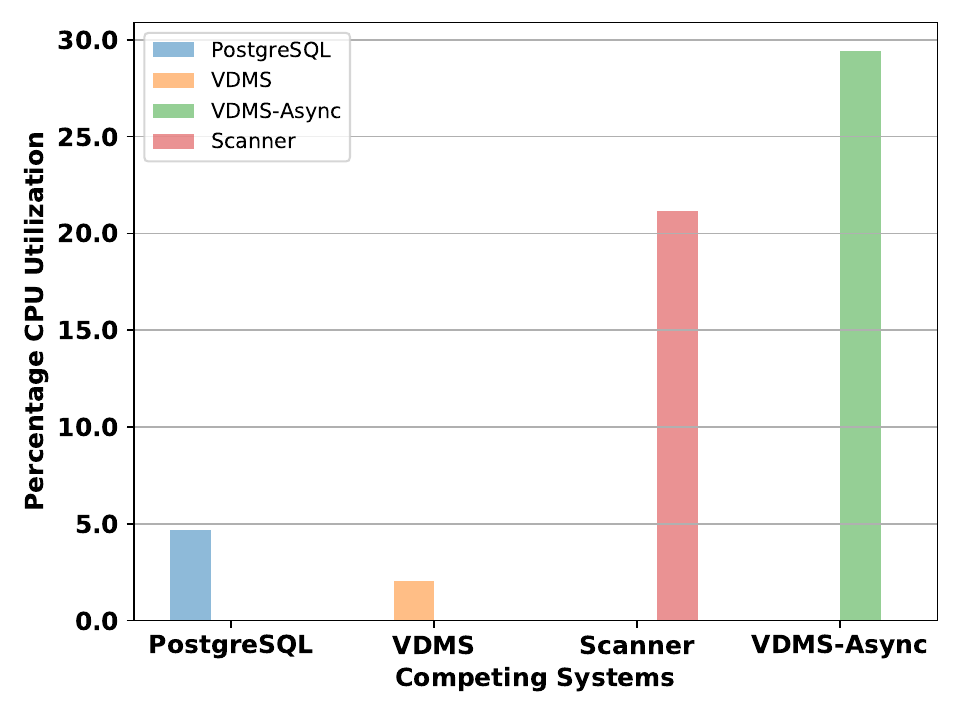}
    \caption{[Video-C2]: Average Resource Usage}
    \label{fig:vcpu_pme}
\end{figure}

\subsubsection{\textbf{Video-C2:} }In the next set of experiments, we query the system with an operation pipeline, Activity Recognition $\rightarrow$ Resize $\rightarrow$ Select $\rightarrow$ Manipulation.

All three plots (Figure~\ref{fig:vpme}-~\ref{fig:vcpu_pme}) follow a similar trend as in C1 for all except Scanner. However, the gains of \our are much higher owing to a more compute-intensive workload. Both PostgreSQL and VDMS spend more time idle-waiting when the operations are being performed. Scanner, on the other hand, is not optimized for operation pipelines. Hence, for every frame, it performs all the operations and stores the resulting frame as a row in the output table. This process is repeated for all frames of all videos, thus increasing the overall execution time. Resource utilization trend is also similar to C1 with low utilization for PostgreSQL and VDMS and comparatively higher utilization for Scanner and \our.

\begin{figure}[!ht]
    \centering
    \includegraphics[width=0.7\linewidth]{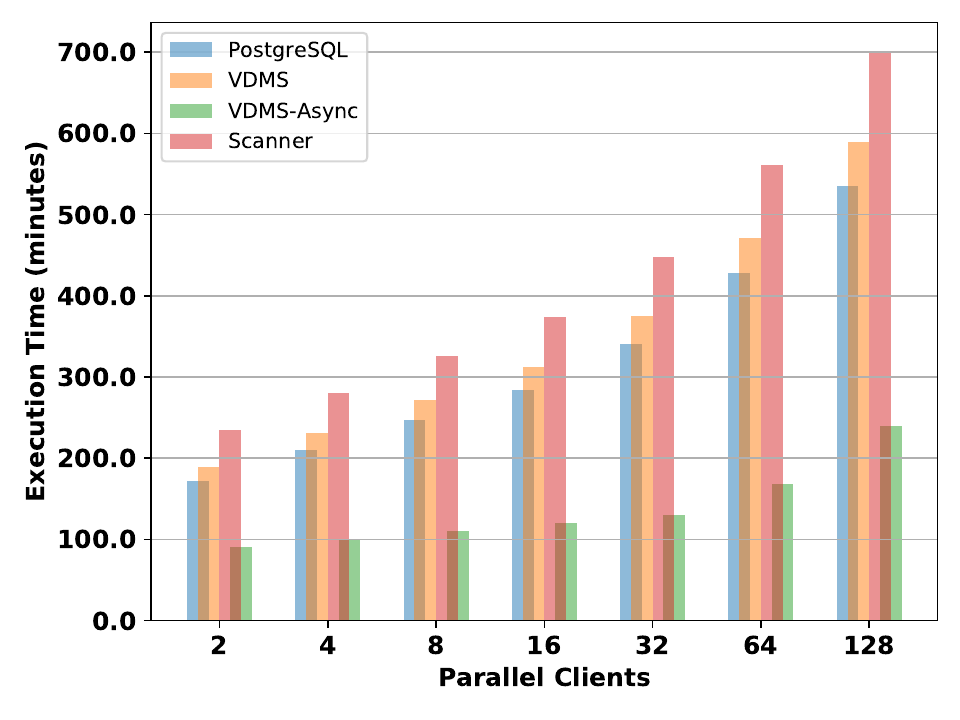}
    \caption{[Video-C3]: Query Duration}
    \label{fig:vmcp}
\end{figure}

\begin{figure}[!ht]
    \centering
    \includegraphics[width=0.7\linewidth]{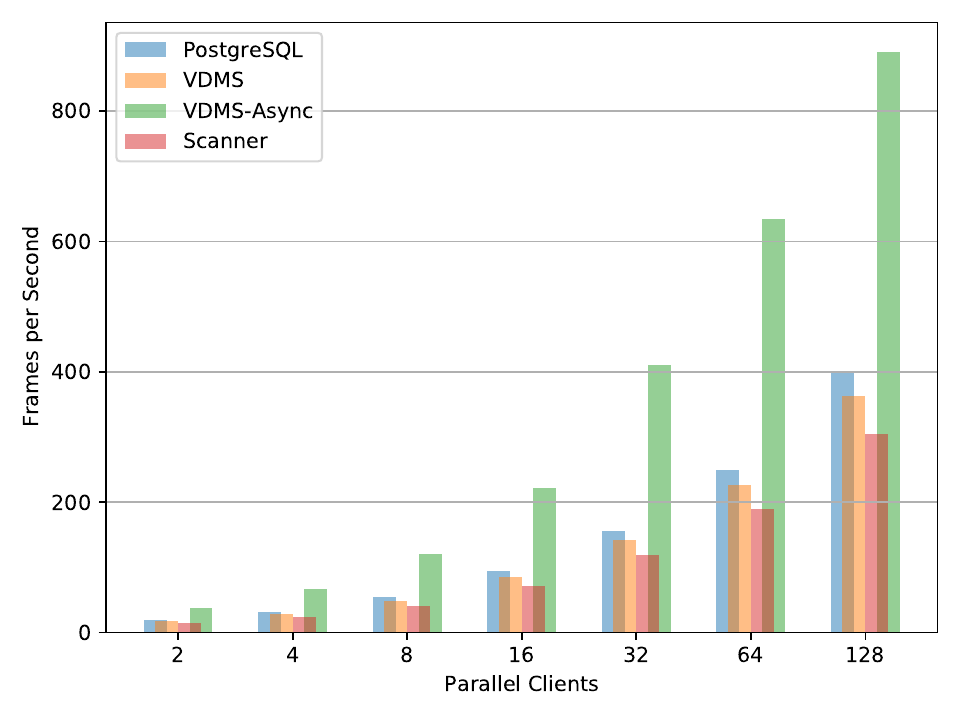}
    \caption{[Video-C3]: Throughput}
    \label{fig:vps_mcp}
\end{figure}

\begin{figure}[!ht]
    \centering
    \includegraphics[width=0.7\linewidth]{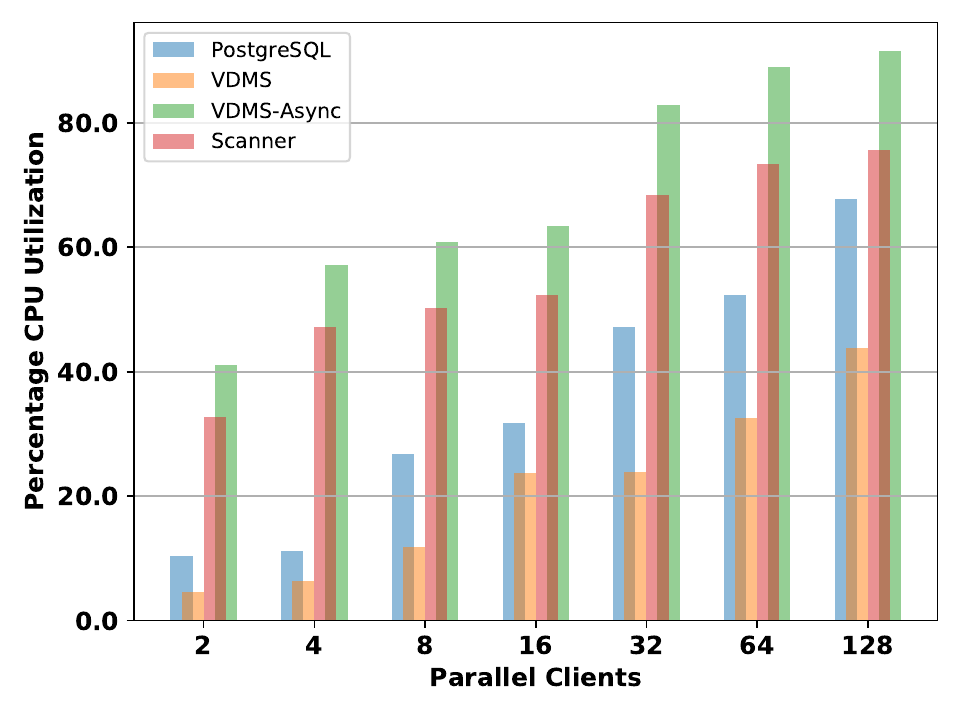}
    \caption{[Video-C3]: Average Resource Usage}
    \label{fig:vcpu_mcp}
\end{figure}

\subsubsection{\textbf{Video-C3:} }We next query the operation pipeline used in C2, but with multiple clients querying the system concurrently. We increase the number of concurrent clients from 2 to 128 by doubling every run.

\our fares better than other systems, and the gain is considerably better compared to the other two categories owing to the higher compute requirement with the parallel clients coming into the picture. For any set of clients, \our has a gain in query duration by at least $2$X of that of the competing systems (Figure~\ref{fig:vmcp}). The gain for the number of frames processed per second goes to $~3$X for the higher number of clients (32, 64) (Figure~\ref{fig:vps_mcp}). However, with the workload becoming much higher for the server with 128 clients, this gain drops to $~2.5$X. The CPU utilization trends (Figure~\ref{fig:vcpu_mcp}) are similar to Image-C3 as with more parallel clients, the CPU usage goes higher for all systems.

\subsection{Result Analysis}
We next analyse the reason behind the results in the benchmarking experiments. For this analysis, we utilize the Visual Road Dataset~\cite{haynes2019visual} that has long videos of $\approx60$ minutes and $\approx200$MB in size and $\approx100$k frames. The longer videos ensure that there is visible idle-waiting time to observe and also show the effectiveness of \our with longer videos. The dataset has 64 different videos, but we perform this analysis with ten videos such that the results are easier to understand. The experiment setup is the same as the previous experiments with the VDMS server, remote server and the client at three different machines. In these experiments, we use VQ7.

\begin{figure}[!ht]
    \centering
    \includegraphics[width=1\linewidth]{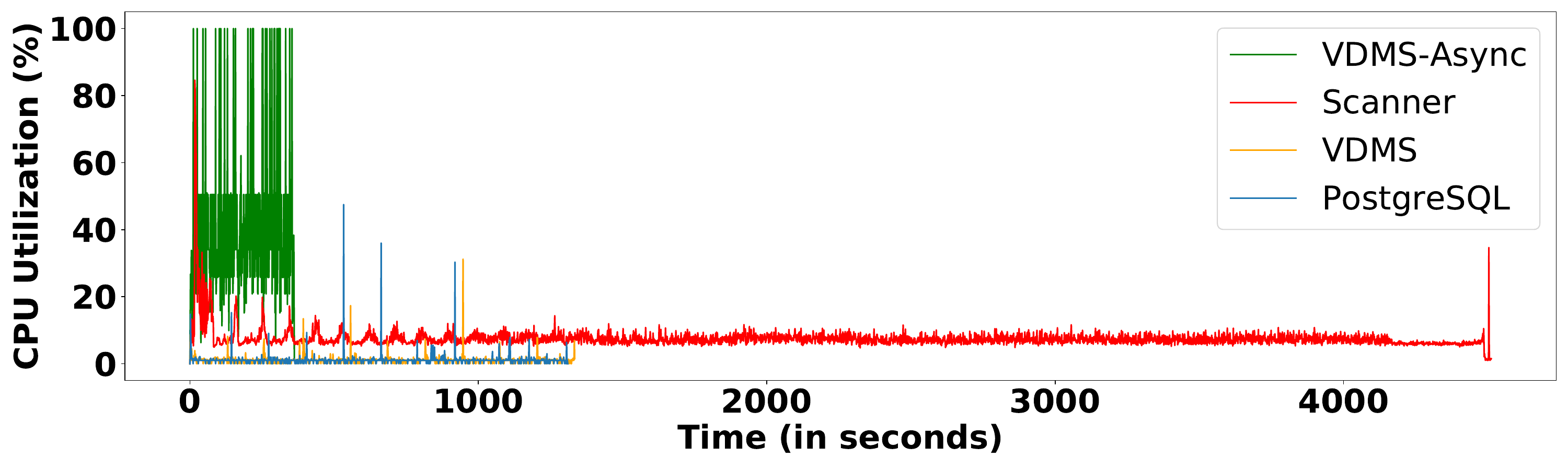}
    \caption{CPU Utilization over time for Competing Systems}
    \label{fig:analysis_cpu}
\end{figure}




\begin{figure}[!ht]
    \centering
    \includegraphics[width=1\linewidth]{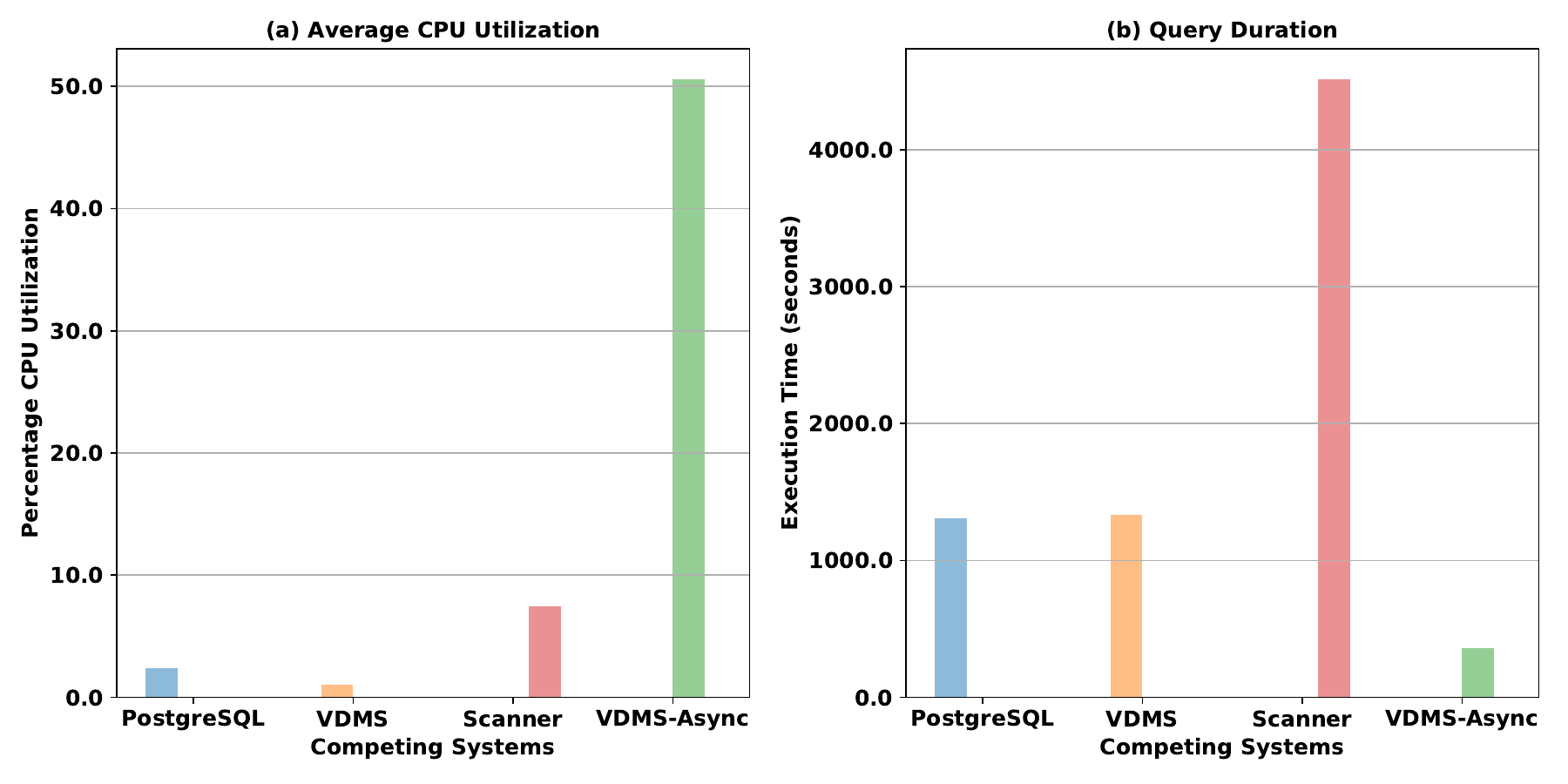}
    \caption{Average CPU Utilization during query execution and Query Duration. \our completes $~\approx3.5X$ faster than PostgreSQL and VDMS, and $\approx12X$ faster than Scanner. }
    \label{fig:analysis_avg}
\end{figure}

Both PostgreSQL and VDMS have periods of idle waiting (Figure~\ref{fig:analysis_cpu}), resulting in longer query duration but very low average CPU utilization (Figure~\ref{fig:analysis_avg}(a)). Scanner works on a frame-by-frame basis with 100k frames resulting in the CPU being almost always active (Figure~\ref{fig:analysis_cpu}). However, the utilization for a single frame is low (indicated by the smaller peaks), leading to overall low CPU Utilization (Figure~\ref{fig:analysis_avg}(a)) but a fairly longer running time. \our completes the entire query with a short burst of high activity (Figure~\ref{fig:analysis_cpu}). A higher level of parallelization with two threads active for the entire query duration leads to a CPU utilization of  $\approx50\%$ on average (Figure~\ref{fig:analysis_avg}(a)). The event-driven architecture of \our eliminates idle waiting and utilizes the available CPU resources to guarantee faster query completion. As seen in Figure~\ref{fig:analysis_avg}(b), there is a 3-12X improvement in query duration against the competing systems. We believe this is a suitable trade-off for a visual data management system that deals with compute-intensive operations.

We note that the higher level of parallelization enabled by \our places a higher load on the remote server(s) and the network. The system-wide effects of such load need to be considered, especially in larger deployments with multiple tenants. Future research in this space could include intelligent offloading with system-wide resource awareness, as well as QoS policies for service differentiation.

\subsection{Scale-out Results}
We next evaluate the impact of scaling out with remote operations. We use a set of AWS Instances to set up an ecosystem of remote servers. All these instances run the Ubuntu 20.04 dual-socket server with AMD EPYC 7R13 CPUs @ 2.65GHz with 192 cores and have 376 GB of DDR4 DRAM. We run our tests on the LFW dataset and execute the query IQ4 on \our. We vary the number of remote servers from 1 to 64 and run two experiments with 32k and 320k images by employing 64 parallel clients querying simultaneously. All results are based on averaging over $15$ runs.

\begin{figure}[!ht]
    \centering
    \includegraphics[width=0.8\linewidth]{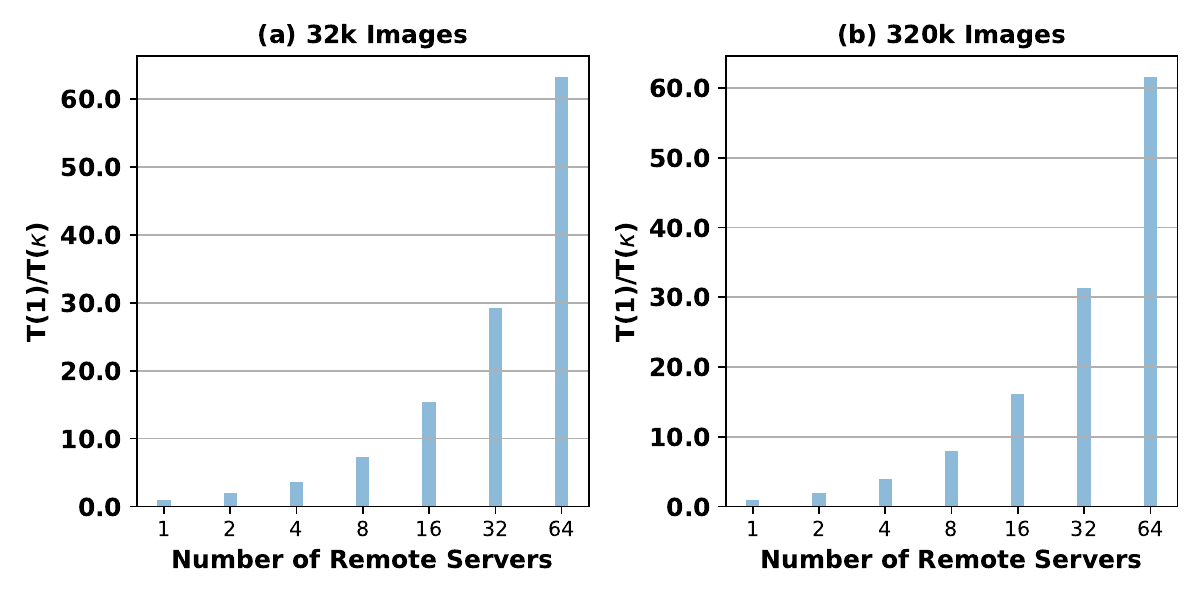}
    \caption{Relative performance gains with remote servers}
    \label{fig:scaleout}
\end{figure}


Figure~\ref{fig:scaleout} shows the ratio of query execution time when one remote server is used to the query execution time when $\kappa$ remote servers are used ($\frac{T(1)}{T(\kappa)}$), where $T(x)$ is the query execution time when $x$ remote servers are used. It is evident that employing $\kappa$ remote servers decreases the execution time by $\kappa$ times. Effectively, increasing the number of remote servers from $1$ to $64$ decreases the query execution time by $~64$X. It should be noted that a limit of 64 servers is kept to show the linear increase in gain, and no scaling limits are reached. The linear reduction will be true for a larger number of servers. The linear scaling of \our opens the possibility for large-scale deployments to significantly reduce overall execution time, or handle large numbers of clients, by simply adding more remote servers.

%% file: Conclusion.tex
\section{Conclusion}\label{conclusion}
The increasing utilization of visual data in a plethora of applications has brought up the need for efficient visual data management systems. However, these systems need to address various challenges like optimizing query execution time, multitasking, end-to-end framework support, supporting a large number of clients, and edge deployments with resource constraints. In light of this, we develop \our, which is an improvement over VDMS, an existing visual data management system. Our upgrade overhauls the existing VDMS architecture to support asynchronous query processing as well as includes support for user-defined operations and remote operations. These capabilities further help scale to a much larger number of clients. Our experiments on \our show that for both image and video data, \our reduces the query execution time by at least half compared to the existing state-of-the-art systems. Moreover, when provided with $\kappa$ remote servers, \our reduces the execution time by $\kappa$ times.

\our thus provides an efficient and fast visual data management system that could be utilized by a large number of applications for both images and videos. There do exist some key directions that require further research. (a)~Designing a distributed version of \our that works on a single data store at a remote location could further increase the scalability. (b)~\our works with videos by extracting all frames, further research is required to devise strategies such that a subset of frames could suffice. (c)~An intelligent setup can be designed that decides when to offload the tasks to a remote server and to what number of remote servers based on the workload and the available resources. 